%% file: main.tex
\documentclass{IEEEtran}
\PassOptionsToPackage{table}{xcolor}

\usepackage{multirow}
\usepackage{balance}
\usepackage{amsmath,amsfonts}
\usepackage{algorithmic}
\usepackage{algorithm}
\usepackage{array}
\usepackage[caption=false,font=normalsize,labelfont=sf,textfont=sf]{subfig}
\usepackage{textcomp}
\usepackage{stfloats}
\usepackage{url}
\usepackage{verbatim}
\usepackage{graphicx}
\usepackage{cite}
\usepackage{comment}
\input{MyPackages}

\begin{document}

\title{Super-Resolution Experimental Validation and Polarimetric Extension of the Effective Roughness Diffuse Scattering Models}
\author{Giacomo Melloni, Jack Chuang, Samuel Berweger, Enrico M. Vitucci, Vittorio Degli-Esposti, Camillo Gentile, and Nada Golmie
\thanks{G. Melloni, J. Chuang, S. Berweger, C. Gentile and N. Golmie are with the Communications Technology Laboratory (CTL), National Institute of Standards and Technology (NIST), Gaithersburg, MD 20899 USA (e-mail: \{first.last\}@nist.gov). G. Melloni is also with the Norwegian University of Science and Technology (NTNU), Trondheim, 7034 Norway.}
\thanks{E. M. Vitucci and V. Degli-Esposti are with the Department of Electrical, Electronic, and Information Engineering “Guglielmo Marconi” (DEI), CNIT, University of Bologna, 40126 Bologna, Italy (e-mail: \{enricomaria.vitucci,v.degliesposti\}@unibo.it).}}



\renewcommand{\Huge}{\fontsize{24}{28}\selectfont}
\maketitle


\begin{abstract}
The experimental validation of diffuse scattering models has long been limited by the inability to spatially separate specular and diffuse contributions in measured channels. This paper overcomes this limitation by combining super-resolution multipath component (MPC) extraction, which resolves individual propagation paths including the specular component, with digital-twin-assisted geometry, enabling the spatial separation of specular and diffuse contributions from bistatic measurements at 28~GHz.
Using this framework, we provide the first measurement-driven validation of the Effective Roughness (ER) model with independent characterization of diffuse scattering across ten common building materials, each measured over 266 angular configurations and all polarization combinations (HH, HV, VH, VV).
Furthermore, we extend the ER framework by proposing a novel angle-dependent cross-polarization discrimination (XPD) model, capturing the geometry-dependent nature of depolarization that is neglected in existing approaches.
The proposed method reproduces the measured diffuse power trends, achieving RMSE values as low as 3 dB across the tested materials, and improves XPD prediction over the baseline constant-XPD model for nearly all material-polarization cases. These results establish a physically consistent and practically viable approach for high-fidelity channel modeling in mmWave systems.
\end{abstract}


\section{Introduction}

Accurate modeling of radio propagation is a fundamental requirement for the design, optimization and simulation of wireless networks. A common simplifying assumption is that building walls and indoor surfaces are flat, smooth, and uniform, under which the Geometrical Optics (GO) approximation \cite{Felsen1973} provides a practical and computationally efficient framework, treating wave interactions as specular reflections, transmissions, and edge diffractions. In practice, however, perfectly smooth surfaces and homogeneous materials are seldom encountered: surface roughness and material inhomogeneities generate a multitude of micro-interactions whose net effect is a redistribution of the incident power in non-specular directions, a phenomenon known as \gls{ds}. The relevance of \gls{ds} has long been underestimated at sub-6 GHz frequencies, where the wavelength is large compared to typical surface and volume irregularities. Next-generation wireless networks, however, will probably migrate toward mm-wave frequencies to exploit wider available bandwidth and meet the exponential growth in capacity demand. At these frequencies, the wavelength becomes comparable to the scale of common surface roughness and material inhomogeneities, making \gls{ds} an even more important propagation mechanism.

Several approaches have been proposed to model \gls{ds}  generated by rough surfaces   \cite{UlabyMooreFung1982_MRS_Vol2, 9759472, Ticconi2011ModelsRoughSurfaces}, including the Kirchhoff approximation (KA), the Small Perturbation Method (SPM), and the Integral Equation Method (IEM). In classical formulations, these models are typically derived under Gaussian surface statistics, which enable tractable analytical or semi-analytical expressions in specific regimes. These methods are widely used in remote sensing applications, particularly for modeling sea surfaces where Gaussian statistical descriptions are commonly adopted. However, such an assumption is hardly verified even for the sea surface \cite{8085399}, not to mention building walls with their geometrical indentations and non-homogeneous compound materials.

\IEEEpubidadjcol

Over the last two decades, the \gls{er} approach has emerged as one of the most popular models for diffuse scattering due to its efficiency and scalability, as demonstrated by extensive research activity  \cite{933491, 4052607, 10137406, melloni2026computationallyefficientreciprocaleffective, 8580765, 8228808, 6813622, Buccioli2021_URSIGASS, 11075837, https://doi.org/10.1155/2020/1583854, 8758806, 7400901, 8288373, 11084884, 9354548}. The \gls{er} model is heuristic and can be applied to arbitrary surfaces, and it has been experimentally validated for indoor and outdoor materials such as brick and drywall. Hence, no prior assumptions about surface statistics are required. In \cite{11084884}, the \gls{er} and KA diffuse scattering models are theoretically compared at 28\,GHz, showing strong agreement. The popularity of the model is supported by its integration in several \gls{rt} tools, including NVIDIA\footnote{\label{fn:disclaimer}Certain commercial equipment, instruments, or materials are identified in this paper to foster understanding. Such identification does not imply recommendation or endorsement by the National Institute of Standards and Technology (NIST).}'s \textit{Sionna RT} \cite{sionna-rt}, Remcom\footnotemark[\getrefnumber{fn:disclaimer}]'s \textit{Wireless InSite} \cite{remcom_wireless_insite}, and Siradel\footnotemark[\getrefnumber{fn:disclaimer}]'s \textit{Volcano} \cite{SIRADEL_VolcanoFlex_2021}.

Nevertheless, further research is required to parameterize the model for different materials \cite{https://doi.org/10.1155/2020/1583854, 8758806, 7400901, 10137406}. In \cite{https://doi.org/10.1155/2020/1583854}, \gls{rf} channel parameters were obtained for several materials at 40, 45, and 50\,GHz, where the scattered power was derived from aggregate power measurements where the reflected, incident, and transmitted contributions were not easily separable. A \gls{rt} calibration method was used to estimate both \gls{sp} and \gls{ds} parameters. The study showed that the predicted received power can increase by up to 20\,dB in some cases when \gls{ds} is included, highlighting its important role in mmWave propagation.

Similarly, \cite{7400901} proposed a parameter estimation procedure for Lambertian and directive \gls{er} diffuse scattering models at 60\,GHz. Several materials were analyzed, including plasterboard walls, cardboard boxes, and brick walls. The extracted parameters were validated through comparison between ray tracing simulations including \gls{er} scattering and measurements, demonstrating the effectiveness of the approach for accurate mmWave channel modeling.

To our knowledge, no prior experimental study has been able to spatially separate \gls{sp} and \gls{ds} for validation and parameterization of the \gls{er} model. The main challenge lies in the need for ultra-high-resolution measurements, which are difficult to realize with standard measurement setups. Nevertheless, such separation is essential to isolate individual mechanisms and derive accurate and independent parameters for each one of them, which would be otherwise impossible.

Furthermore, limited understanding exists regarding the impact of \gls{ds} on depolarization \cite{8580765, 6813622, 4458651, 5979177}. In \cite{6183484}, a model was proposed that separates the total \gls{ds} power into \gls{cp} and \gls{xp} components. The distribution of power between the two polarization states is described by the \textit{depolarization factor}, which acts as a scaling factor. In dense \gls{mpc} environments, this model has proven effective in predicting \gls{xpd}. However, depolarization may depend on both observation and incidence angles. Therefore, a spatially aware \gls{xpd} model is needed, yet such a model is currently missing in the literature.

In this work, our main contributions are threefold:

\begin{enumerate}
\item Extensive validation of the \gls{er} modeling approach is conducted for the diffuse scattering channel. The model is evaluated using ten common building materials, yielding ten datasets with a total of 266 distinct observations. Furthermore, the model is validated across all transmitter and receiver polarization combinations.
\item We propose a method that, for the first time, resolves the \gls{sp} and \gls{ds} components in the spatial domain using only measured scattering data, thereby enabling the validation of different directive diffuse scattering models;
\item We extend existing \gls{xpd} modeling by introducing a novel, spatially aware \gls{xpd} model based on \gls{er} theory, which is validated using the full-polarimetric dataset for all ten materials.
\end{enumerate}

The remainder of this paper is organized as follows. Section \ref{sec:background} summarizes the theoretical tools used in this work, including \gls{er} theory and depolarization concepts. Section \ref{sec:xpd} introduces the proposed \gls{xpd} model and its motivation. Section \ref{sec:exp_val} presents the experimental validation methodology and discusses the results. Finally, Section \ref{sec:conclusion} summarizes the main findings.

\section{Background Theory}\label{sec:background}
This section summarizes the theoretical background required for the validation and extension of the effective roughness models. The original effective roughness model (ER) and its reciprocity-compliant extensions (RER and G-RER) are first reviewed, followed by their baseline polarimetric model, which serves as the benchmark for the proposed angle-dependent extension in the next section.

\subsection{The Effective Roughness Models}

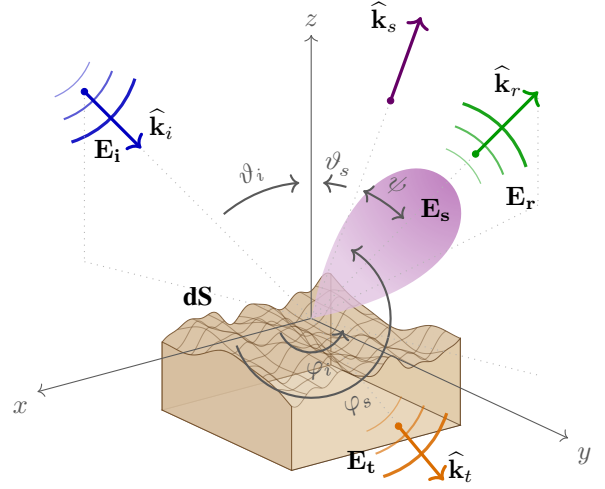
\begin{figure}[!t]
\centering
\input{Pictures/ReferenceSystem}
\caption{Illustration of the \gls{em} fields under consideration, i.e., the incident, reflected, scattered, and transmitted fields ($\mathbf{E_i}$, $\mathbf{E_r}$, $\mathbf{E_s}$, $\mathbf{E_t}$).\label{fig:ReferenceSystem}}
\vspace{-0.5cm}
\end{figure}

The effective roughness framework for modeling \gls{ds} components assumes that diffuse scattering power is redistributed in all directions at the expense of specular reflection, thereby enforcing a physically consistent power balance.
This section reviews the evolution of the framework, from the ER model to its reciprocity-compliant extensions, namely the \gls{rer} and \gls{g-rer} models.

Let $\hat{\mathbf{k}}_i$ and $\hat{\mathbf{k}}_s$ denote the unit wave vectors associated with the incident and scattered field, which depend on the spherical angles $(\vartheta_i,\varphi_i)$ and $(\vartheta_s,\varphi_s)$ as depicted in Fig.\,\ref{fig:ReferenceSystem}. The figure represents the pattern of the \gls{ds} field $\mathbf{E}_s$ scattered in non-specular directions, as well as the reflected and transmitted fields $\mathbf{E}_r, \mathbf{E}_t$ with their respective directions $\hat{\mathbf{k}}_r,\hat{\mathbf{k}}_t$, computed through the geometrical optics theory.

The intensity of the diffuse scattering field $\lvert \mathbf{E}_\text{S} \rvert^{2}$ radiated by the surface element $\mathrm{dS}$ is given by \cite{10137406}
\begin{equation}
\lvert \mathbf{E}_\text{S} \rvert ^{2}
= \left( \frac{K_{\scriptscriptstyle \text{TX}}\, S}{r_i\, r_s} \right)^{2}
\, \Gamma^{2}(\vartheta_i)\, \cos(\vartheta_i)\,
\frac{f(\hat{\mathbf{k}}_s , \hat{\mathbf{k}}_i)}{F(\hat{\mathbf{k}}_i)}\, \mathrm{dS} \, ,
\label{eq:e_s}
\end{equation}

In \eqref{eq:e_s}, the factor $K_{\scriptscriptstyle \text{TX}}$ accounts for the transmit power and antenna gain, $S\in \left[ {0,1} \right]$ is the \textit{scattering parameter}, with $S^2$ expressing the percentage of power diffused in non-specular directions at the expense of specular reflection, $f(\hat{\mathbf{k}}_s,\hat{\mathbf{k}}_i)\equiv f(\vartheta_i,\varphi_i,\vartheta_s,\varphi_s)$ represents the \textit{scattering pattern}, and $r_i$, $r_s$ denote the propagation distances from the \gls{tx} to $\mathrm{dS}$ and from $\mathrm{dS}$ to the \gls{rx}, respectively.
Moreover, $\vartheta_i=\arccos{(-{{\mathbf{\hat k}}}_i \cdot {\mathbf{\hat z}})}$ is the incidence angle, with ${\mathbf{\hat z}}$ standing for the unit vector orthogonal to the surface, and $\Gamma^2(\vartheta_i)=|\mathbf{E}_r|^2/|\mathbf{E}_i|^2$ is the power reflectivity coefficient, computed through the standard Fresnel's theory.

Finally, $F(\hat{\mathbf{k}}_i)$ is a proper \textit{normalization factor} ensuring that the power balance is satisfied, and computed by integrating the scattering pattern $f$ on the back-scattering half-space with respect to the variables $(\vartheta_s,\varphi_s)$:

\begin{equation}\label{eq:F}
F(\hat{\mathbf{k}}_i)=\int_{0}^{2\pi}\!\int_{0}^{\pi/2}
f(\vartheta_i,\varphi_i,\vartheta_s,\varphi_s)\,\sin(\vartheta_s)\,\dd\vartheta_s\,\dd\varphi_s \, .
\end{equation}

\vspace{0.2em}
\subsubsection{The ER Model}

The ER directive model assumes a single diffuse scattering lobe centered on the specular reflection direction \cite{4052607}
\begin{equation}
f_{\scriptscriptstyle \text{ER}}(\hat{\mathbf{k}}_s , \hat{\mathbf{k}}_i)
= \left( \frac{1 + \cos(\psi)}{2} \right)^{\alpha_{\scriptscriptstyle R}},
\end{equation}
where $\alpha_{\scriptscriptstyle R}$ controls the lobe directivity (the higher $\alpha_{\scriptscriptstyle R}$, the narrower the scattering lobe) and the \textit{off-specular angle} $\psi$ is defined as
\begin{equation}
\label{eq:off-spec}
\cos(\psi)={{{\mathbf{\hat k}}}_s \cdot {{\mathbf{\hat k}}}_r}={\mathbf{\hat k}_s} \cdot \left[ {{{{\mathbf{\hat k}}}_i} - 2( {{{{\mathbf{\hat k}}}_i} \cdot {\mathbf{\hat z}}}){\mathbf{\hat z}}} \right]
\end{equation}
i.e., the angle between the specular reflection and the observation directions. While this model captures the directional spreading of diffuse scattering around the specular direction, it does not satisfy reciprocity, i.e., invariance of the scattering response when the incident and observation directions are interchanged \cite{VanBladel2007}.

\vspace{0.2em}
\subsubsection{The RER Model}

To ensure physical consistency, the reciprocity condition must be satisfied, which can be expressed as
\begin{equation}
\frac{f(\hat{\mathbf{k}}_s , \hat{\mathbf{k}}_i)}{F(\hat{\mathbf{k}}_i)} \cdot \cos(\vartheta_i)
=
\frac{f(\hat{\mathbf{k}}_i , \hat{\mathbf{k}}_s)}{F(\hat{\mathbf{k}}_s)} \cdot \cos(\vartheta_s).
\end{equation}

The reciprocal effective roughness (RER) model satisfies this condition by modifying the directive pattern as \cite{10137406}
\begin{equation}
f_{\scriptscriptstyle \text{RER}}(\hat{\mathbf{k}}_s , \hat{\mathbf{k}}_i)
= \sqrt{\cos(\vartheta_s)} \cdot \left( \frac{1 + \cos(\psi)}{2} \right)^{\alpha_{\scriptscriptstyle R}}.
\end{equation}
This formulation preserves the ER lobe structure while ensuring reciprocity through the inclusion of the $\sqrt{\cos(\vartheta_s)}$ factor.

\subsubsection{The G-RER Model}
g
A further refinement of the RER model is obtained by replacing the polynomial lobe with a Gaussian-shaped directive pattern \cite{melloni2026computationallyefficientreciprocaleffective}
\begin{equation}\label{eq:gaussianModelrec}
\begin{aligned}
    f_{\scriptscriptstyle \text{G\mbox{-}RER}}(\hat{\mathbf{k}}_s , \hat{\mathbf{k}}_i) &= \sqrt{\cos(\vartheta_s)} \cdot \exp \{-\sfrac{1}{2} \; \alpha_{\scriptscriptstyle R} || \hat{\mathbf{k}}_r - \hat{\mathbf{k}}_s ||^2\} \\
    &= \sqrt{\cos(\vartheta_s)} \cdot \exp \{- \alpha_{\scriptscriptstyle R} [1 - \cos(\psi)]\}
\end{aligned}
\end{equation}
This formulation retains the reciprocity property while providing a more efficient representation of highly directive scattering lobes. In particular, the Gaussian form allows achieving comparable directivity with smaller values of $\alpha_{\scriptscriptstyle R}$ and enables more efficient computation of the normalization factor.
The normalization factors for both RER and G-RER admit closed-form or accurate approximations \cite{10137406, melloni2026computationallyefficientreciprocaleffective}.

Due to their comparable predictive performance, and given the improved physical consistency and computational efficiency of the G-RER formulation, it is adopted as the reference model for all three in the remainder of this work.

\subsection{Baseline XPD Model}
\label{subsec:angle-independent}

Depolarization describes the conversion of an incident electromagnetic field into orthogonal polarization components due to interaction with a surface. In scattering-based channel models, this effect is typically represented by decomposing the received field into co-polarized (CP) and cross-polarized (XP) components \cite{Tsang2000ScatteringOE}
\begin{equation}\label{eq:e_tot}
    \mathbf{E_{\scriptscriptstyle \text{RX}}} = \mathbf{E_{\scriptscriptstyle \text{CP}}} + \mathbf{E_{\scriptscriptstyle \text{XP}}} \, ,
\end{equation}
where $\mathbf{E_{\scriptscriptstyle \text{CP}}}$ and $\mathbf{E_{\scriptscriptstyle \text{XP}}}$ denote the CP and XP fields, respectively. The CP component is obtained by projecting the total field at the receiver onto the polarization vector $\hat{\mathbf{e}}_{\scriptscriptstyle \mathrm{p}}$, aligned with the transmit polarization, while the XP component is obtained by projection onto the orthogonal vector $\hat{\mathbf{e}}_{\scriptscriptstyle \mathrm{q}}$.

In previous works, it was shown that \gls{ds} has a prevalent role in the generation of field depolarization, compared to coherent propagation mechanisms (specular reflection, diffraction) that generally preserve the original polarization, except in very particular cases \cite{4458651, 5979177}. This happens due to complex interactions of the \gls{em} signal with rough surfaces, such as irregular building walls, sea surface, etc. \cite{Tsang2000ScatteringOE}

Focusing on the \gls{ds} contribution, \eqref{eq:e_tot} can be expressed as proposed in \cite{6183484}
\begin{equation}\label{eq:e_tot_ds_vitucci}
\begin{aligned}
    \mathbf{E_{\scriptscriptstyle \mathrm{S}}}
    &= E_{\scriptscriptstyle \mathrm{S,pp}} \hat{\mathbf{e}}_{\scriptscriptstyle \mathrm{p}} + E_{\scriptscriptstyle \mathrm{S,pq}}\hat{\mathbf{e}}_{\scriptscriptstyle \mathrm{q}} \\
    &= \sqrt{1-\kappa_p} \; |\mathbf{E_{\scriptscriptstyle \mathrm{S}}}| \, e^{-j \beta_{pp}}\; \hat{\mathbf{e}}_{\scriptscriptstyle \mathrm{p}}
    + \sqrt{\kappa_p} \; |\mathbf{E_{\scriptscriptstyle \mathrm{S}}}| \; e^{-j \beta_{pq}} \; \hat{\mathbf{e}}_{\scriptscriptstyle \mathrm{q}} \, ,
\end{aligned}
\end{equation}
where $|\mathbf{E_{\scriptscriptstyle \mathrm{S}}}|$ is given by \eqref{eq:e_s}, $p,q \in \{\mathrm{H}, \mathrm{V}\}$, with $p$ being the polarization of the field generated by the \gls{tx} antenna and incident on the scattering surface, and $q$ being the polarization orthogonal to $p$, while $\kappa_p$ is the \textit{depolarization factor} that sets the power split between the \gls{cp} and \gls{xp} components, and $\beta_{pp}$, $\beta_{pq}$ $\in \left[-\pi,\pi\right]$ are independent and uniformly distributed random phases.
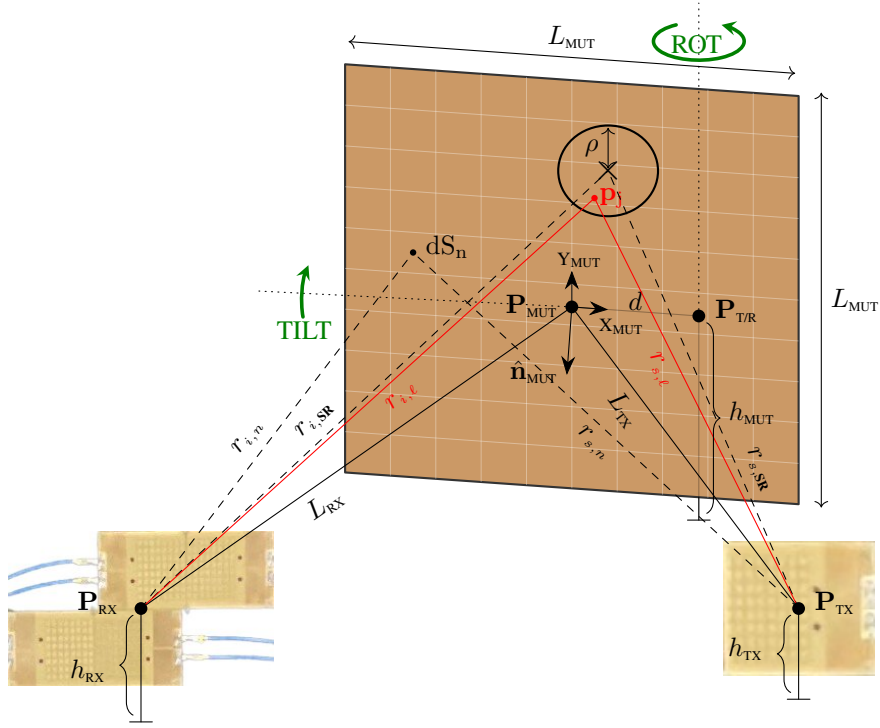
\begin{figure*}[!t]
\centering
\input{Pictures/Testbed}
\caption{Schematic of the measurement procedure. Distances $L_{\scriptscriptstyle \text{RX}}$ and $L_{\scriptscriptstyle \text{TX}}$ are referenced to the antenna centers and the MUT center when the rotator tilt is $0^\circ$ and rotation is $30^\circ$, giving a $60^\circ$ angle between RX and TX pointing directions. The planar surface is discretized into tiles to compute the predicted total \gls{ds} power. An example specular reflection cluster region is also shown. The point $\mathbf{p}_\mathrm{j}$ represents an arbitrary measured \gls{mpc} located within the circle, which will be identified as a specular reflection \gls{mpc}.\label{fig:testbed}}
\end{figure*}
The formulation in \cite{6183484} assumes that the depolarization factor $\kappa_p$ is independent of the incidence/observation directions, i.e., of the scattering geometry.

The corresponding cross-polarization discrimination (XPD) evaluated at \gls{rx} side is computed as the ratio between \gls{cp} and \gls{xp} powers, i.e.
\begin{equation}\label{eq:xpd_vitucci}
    \mathrm{XPD}_p = \frac{P_{\scriptscriptstyle S,\text{CP}}}{P_{\scriptscriptstyle S,\text{XP}}}=\frac{\mathbb{E}\left[|E_{\scriptscriptstyle \mathrm{S,pp}}|^2\right]}{\mathbb{E}\left[|E_{\scriptscriptstyle \mathrm{S,pq}}|^2\right]}
    = \frac{1 - \kappa_p}{\kappa_p} \, .
\end{equation}

This model relies on two key assumptions:

\begin{enumerate}
    \item The depolarization effect is represented by a single scalar parameter $\kappa_p$, which is independent of the scattering geometry;
    \item The CP and XP diffuse components share the same angular behavior. Within the ER framework, this implies that the directive pattern parameters (e.g., $\alpha_{\scriptscriptstyle R}$ in the ER, RER, and G-RER models) are identical for the two polarization components.
\end{enumerate}

\section{Angle-Dependent XPD Model}\label{sec:xpd}

\begin{figure*}[t]
\centering
\includegraphics[width=\linewidth]{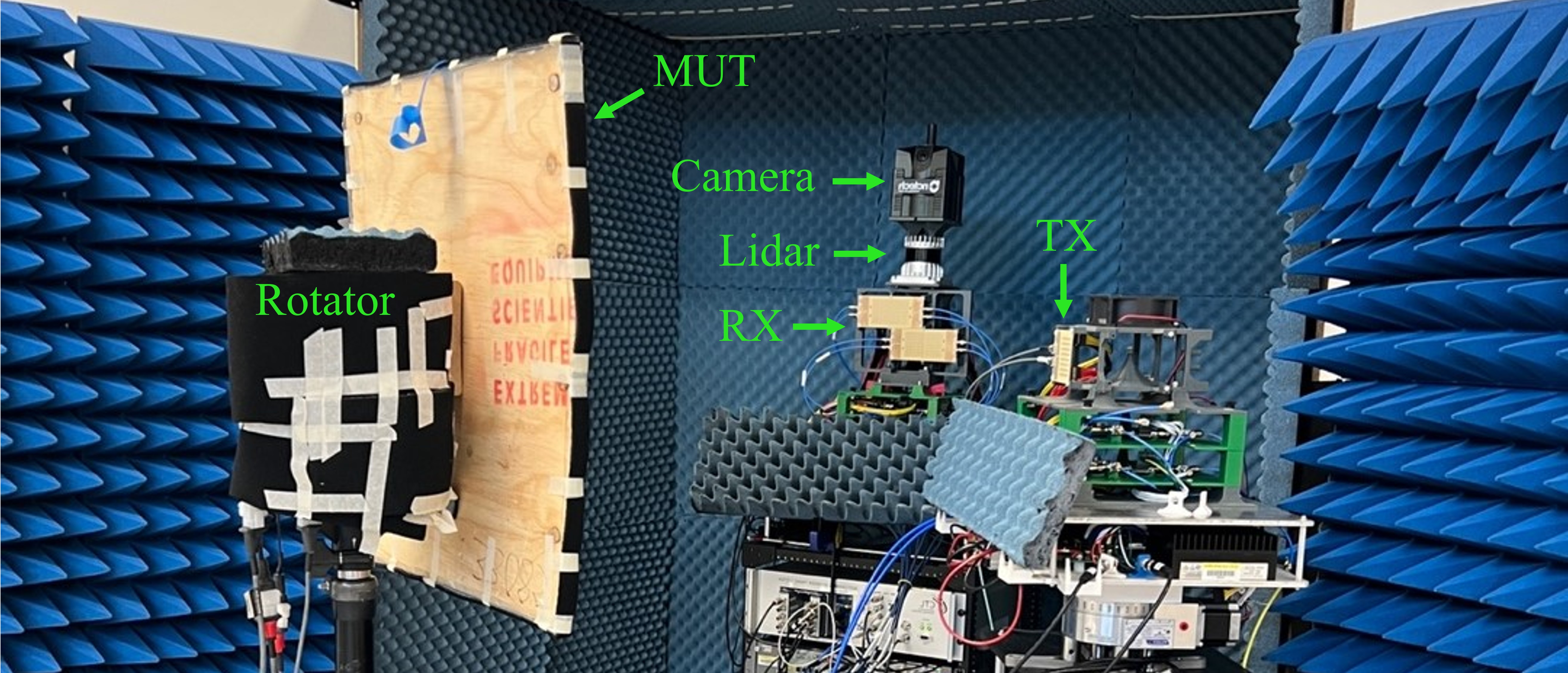}
\caption{Context-aware channel sounder positioned in a semi-anechoic chamber facing the mounted \gls{mut}. From \cite{Sloane2025XPD}.}
\label{fig:setup_photo}
\end{figure*}

\begin{figure*}[t]
\centering
\subfloat[Brick\label{fig:mat_brick}]{%
  \includegraphics[width=0.19\textwidth]{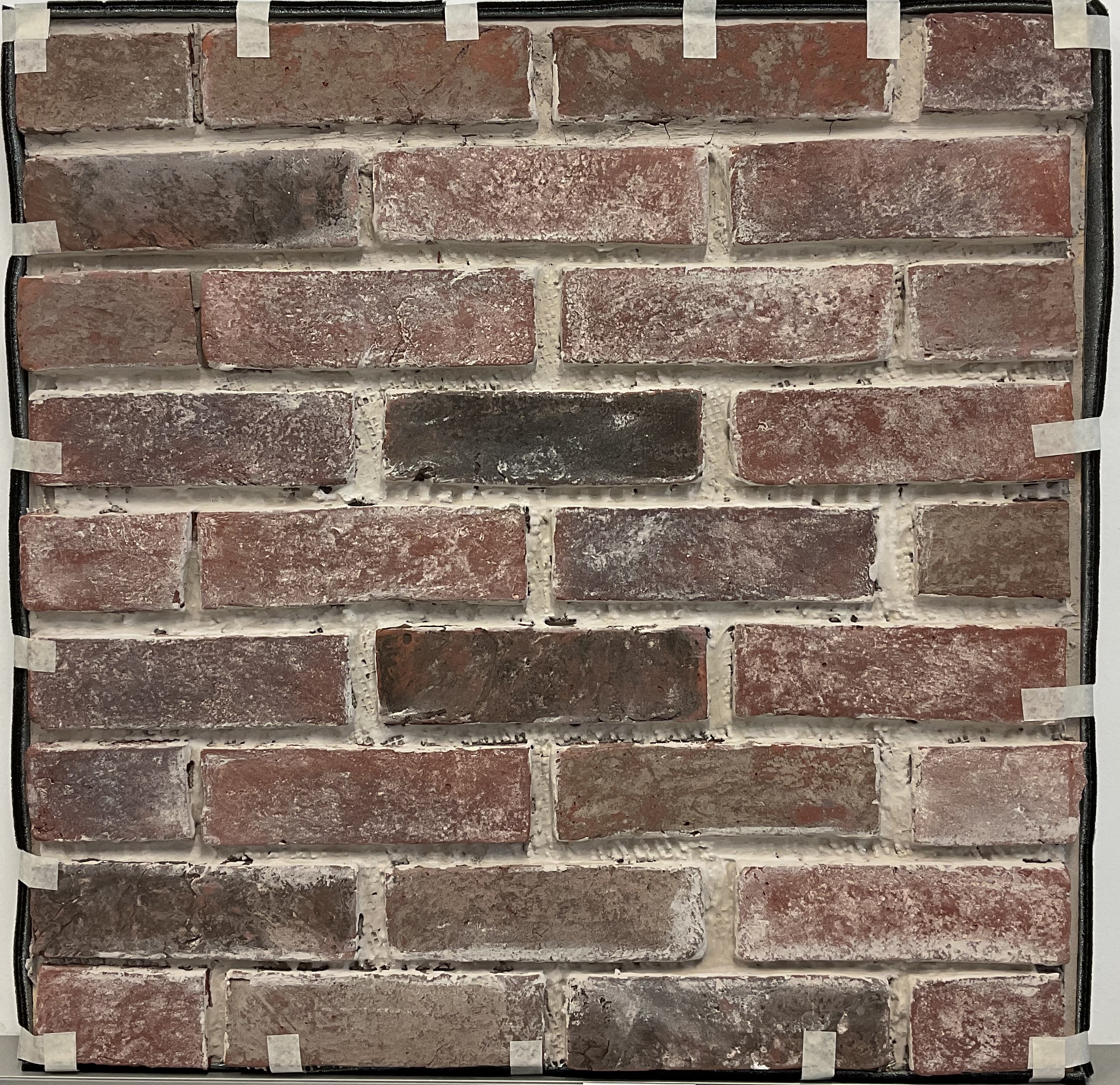}%
}\hfill
\subfloat[Shingles\label{fig:mat_shingles}]{%
  \includegraphics[width=0.19\textwidth]{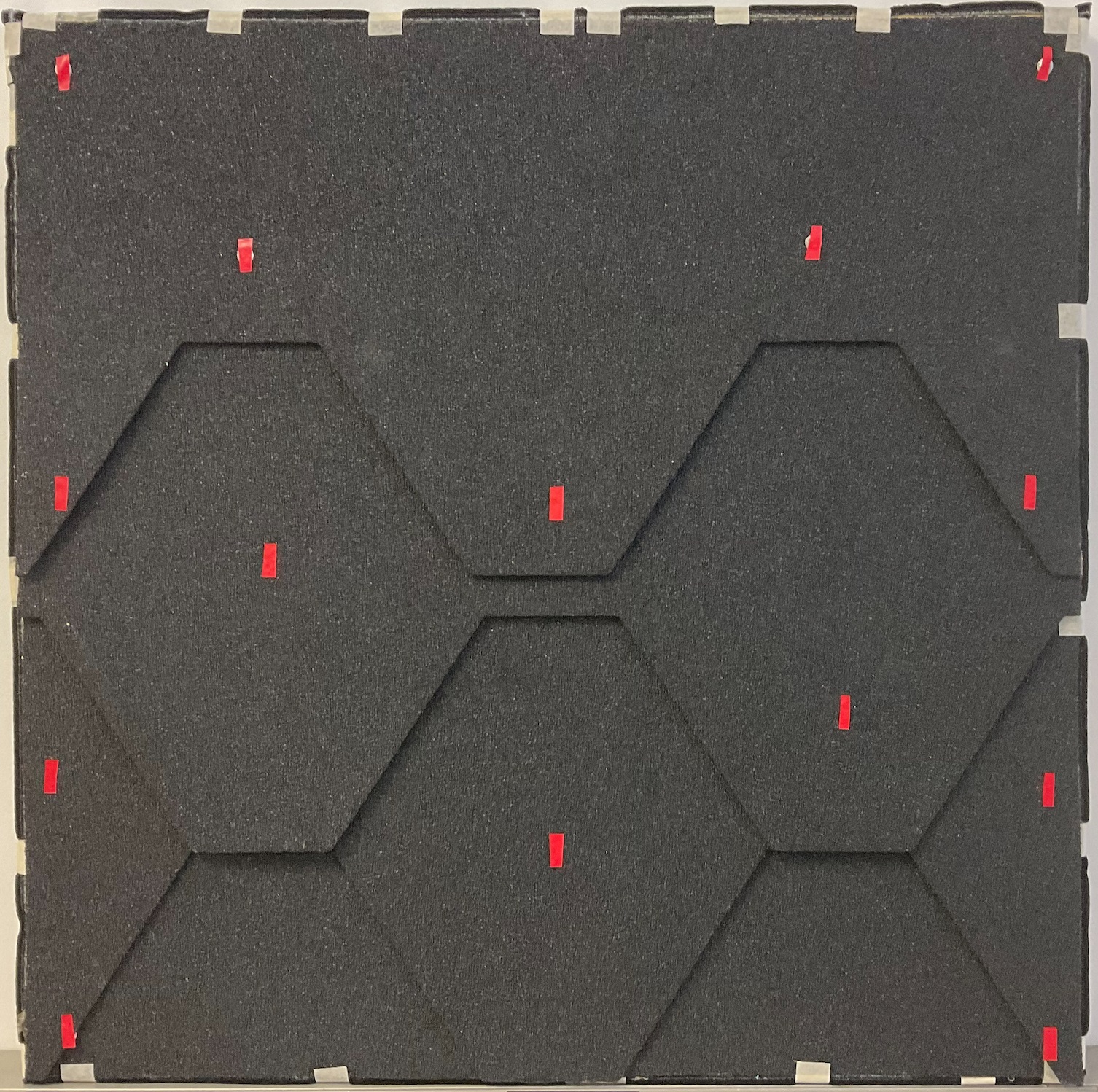}%
}\hfill
\subfloat[Tile\label{fig:mat_tile}]{%
  \includegraphics[width=0.19\textwidth]{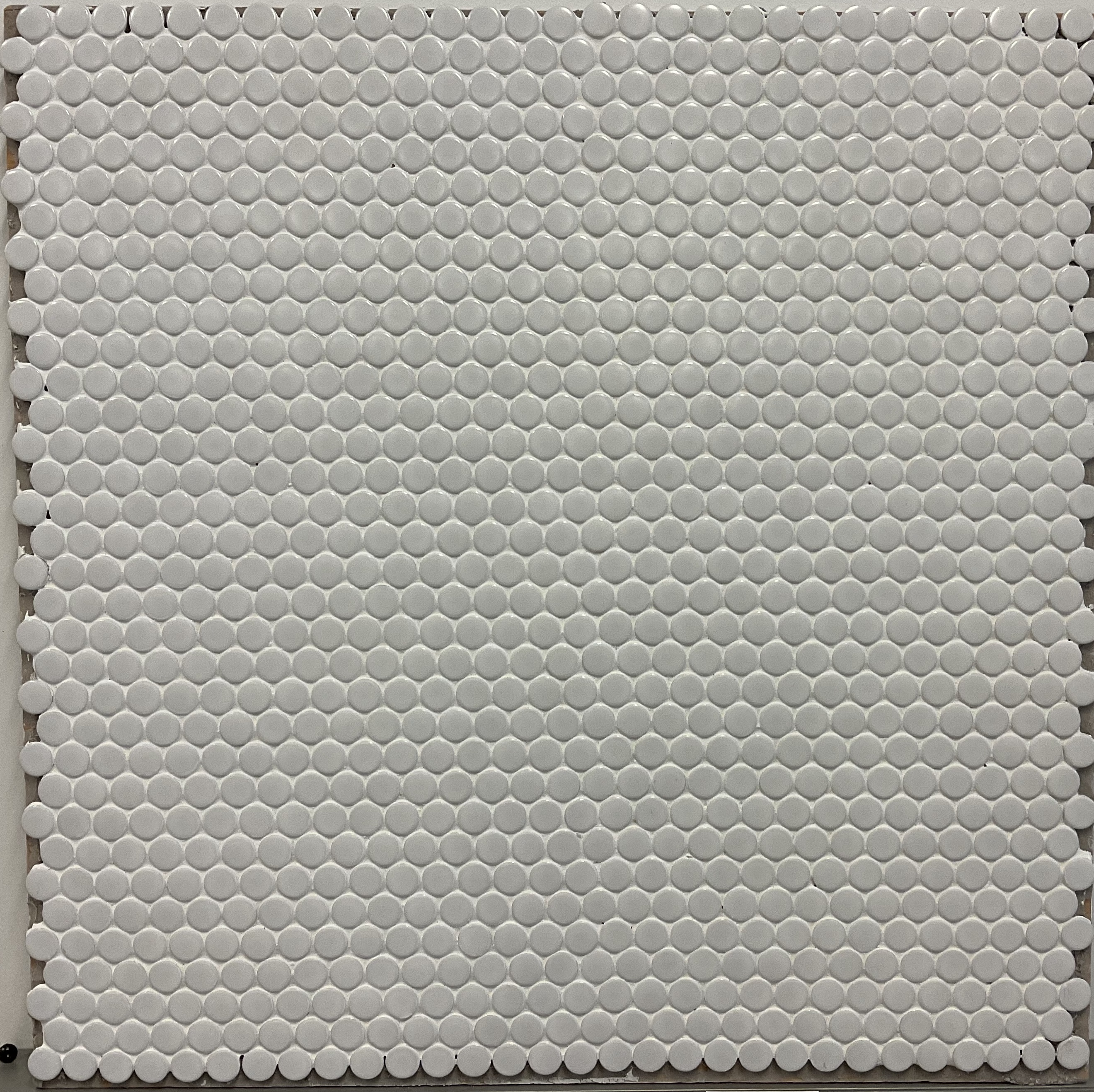}%
}\hfill
\subfloat[Carpet\label{fig:mat_carpet}]{%
  \includegraphics[width=0.19\textwidth]{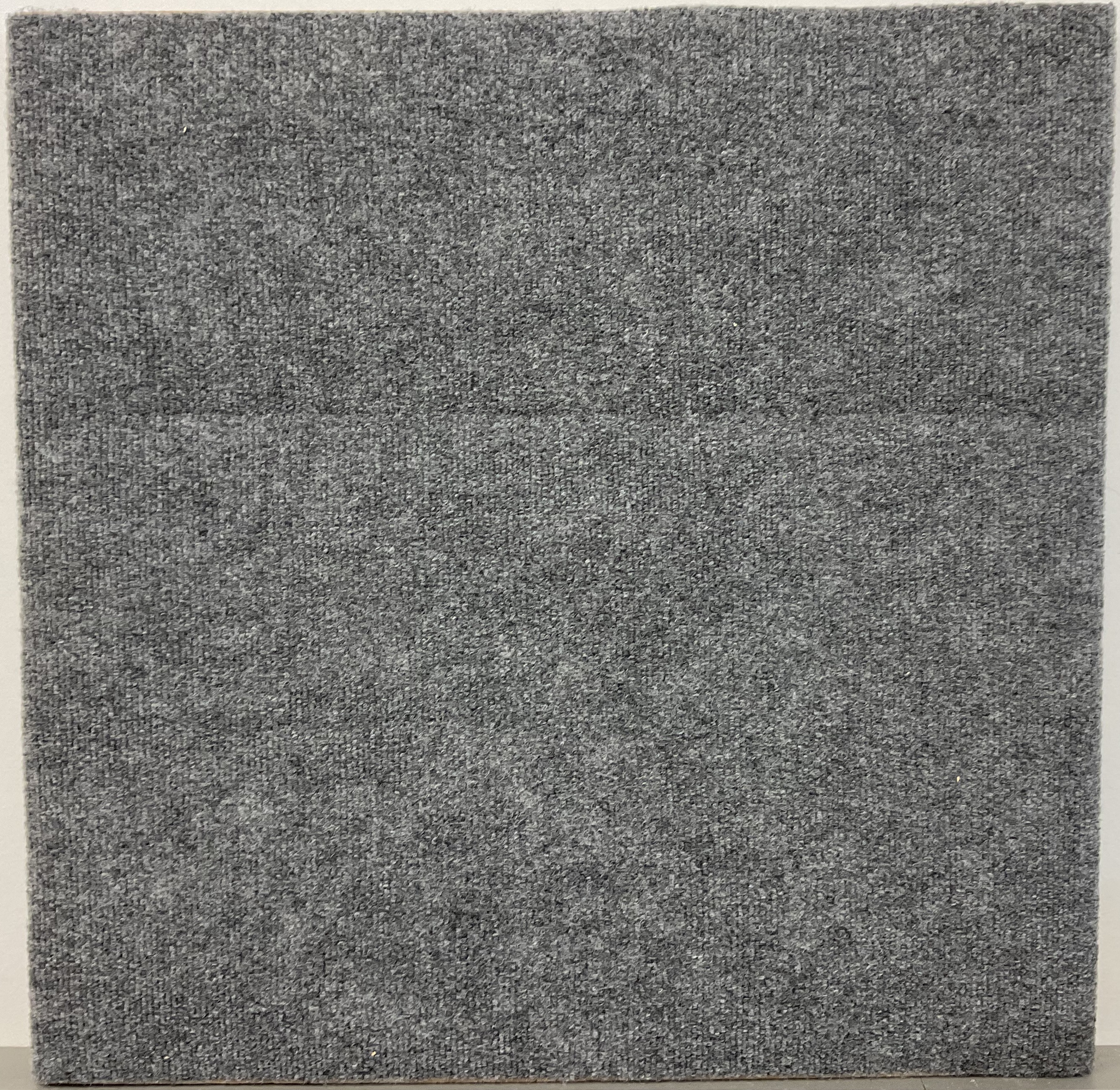}%
}\hfill
\subfloat[Drywall\label{fig:mat_drywall}]{%
  \includegraphics[width=0.19\textwidth]{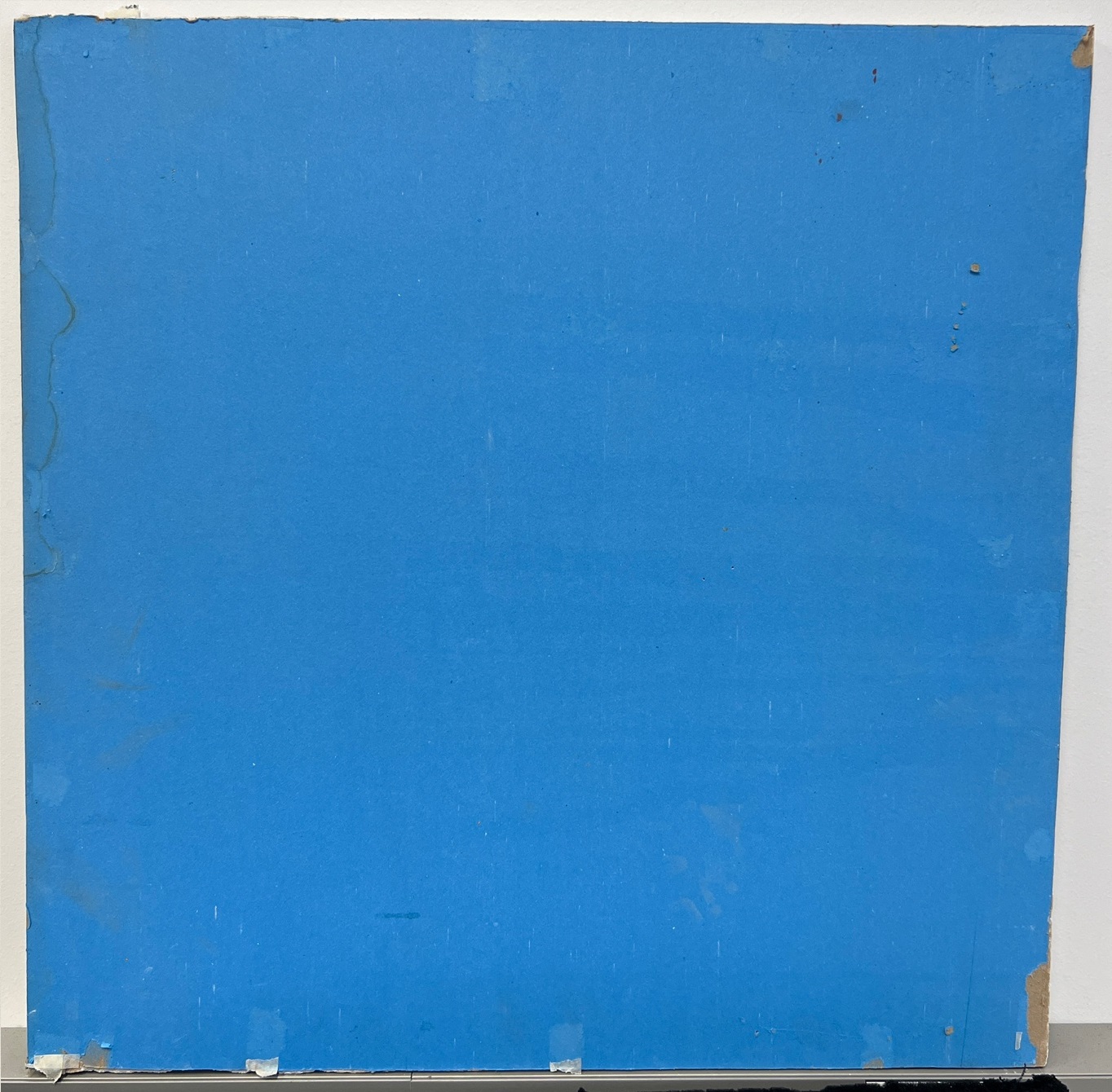}%
}
\vspace{0.5em}
\subfloat[Mortar\label{fig:mat_mortar}]{%
  \includegraphics[width=0.19\textwidth]{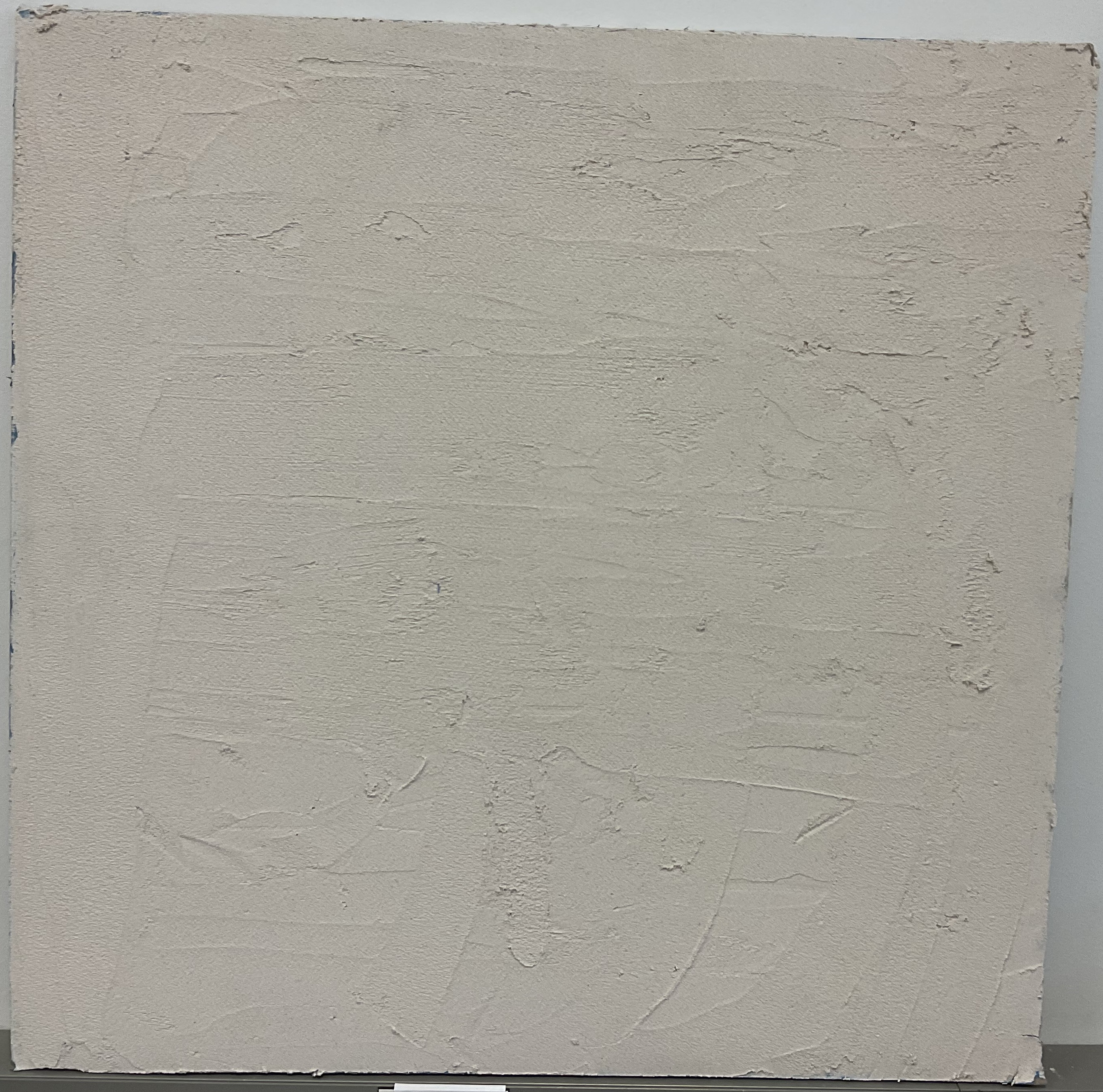}%
}\hfill
\subfloat[Plexiglass\label{fig:mat_plexi}]{%
  \includegraphics[width=0.19\textwidth]{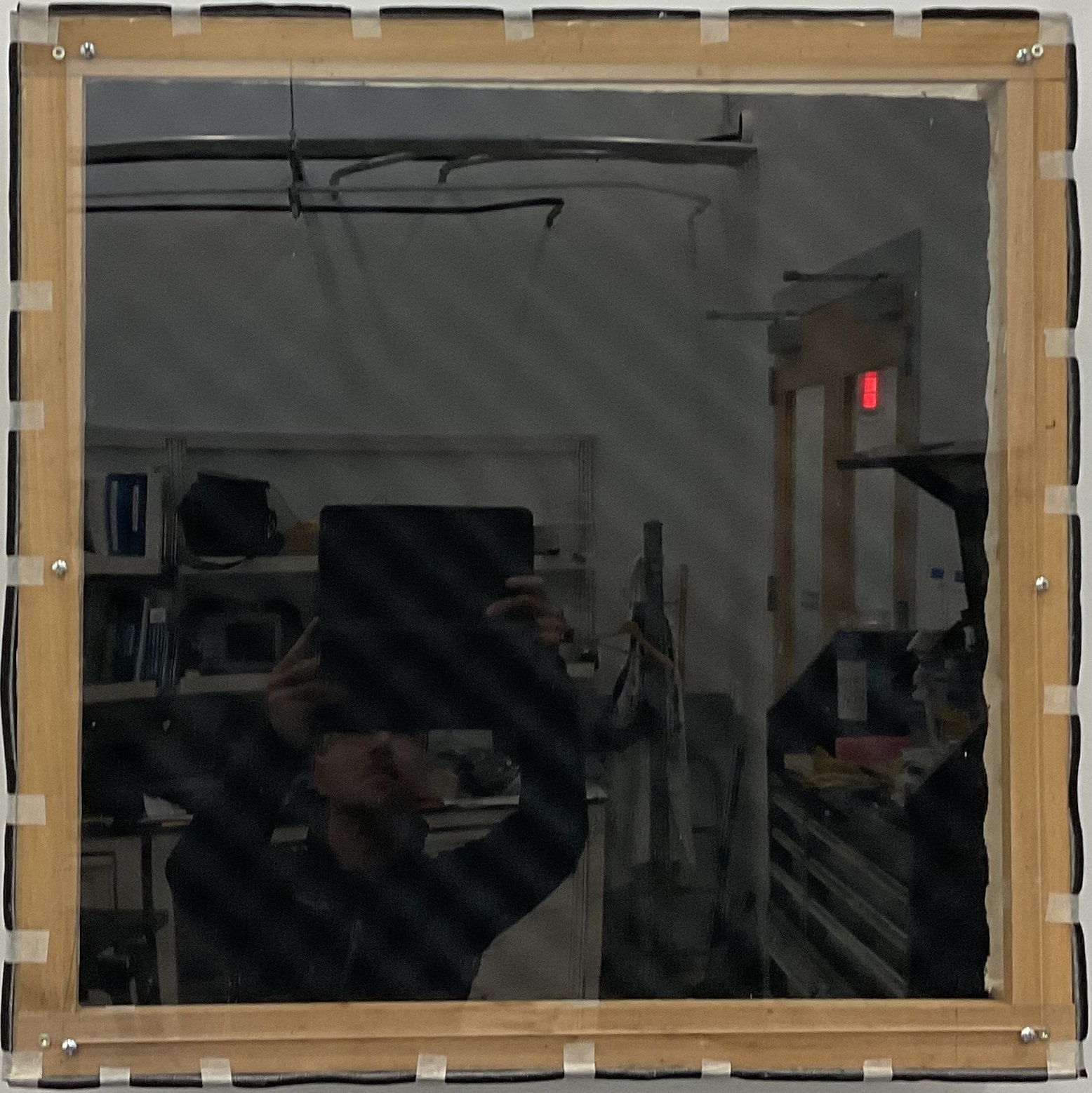}%
}\hfill
\subfloat[Plywood\label{fig:mat_wood}]{%
  \includegraphics[width=0.19\textwidth]{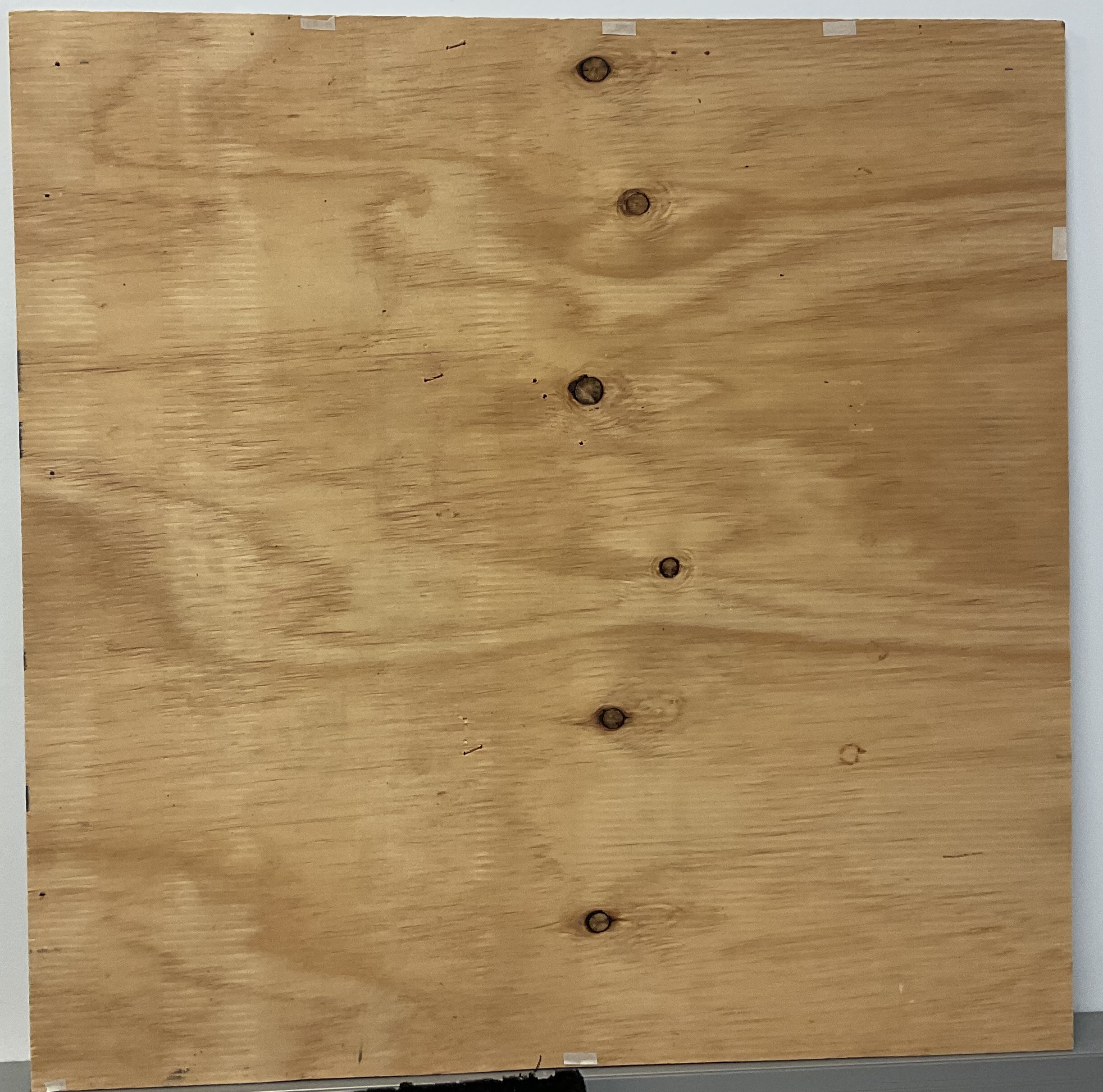}%
}\hfill
\subfloat[Fence 45$^\circ$\label{fig:mat_fence45}]{%
  \includegraphics[width=0.19\textwidth]{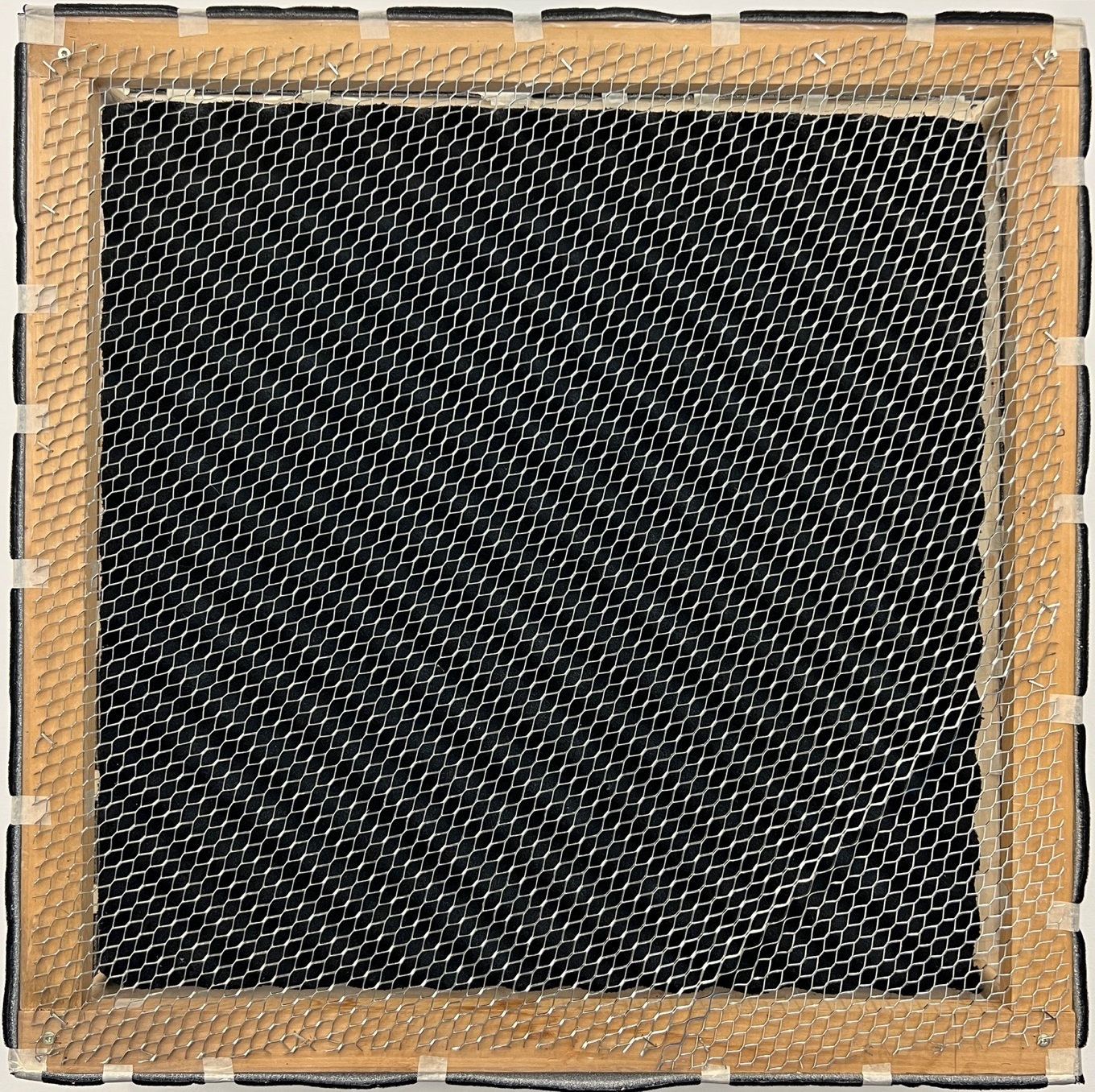}%
}\hfill
\subfloat[Fence 0$^\circ$\label{fig:mat_fence0}]{%
  \includegraphics[width=0.19\textwidth]{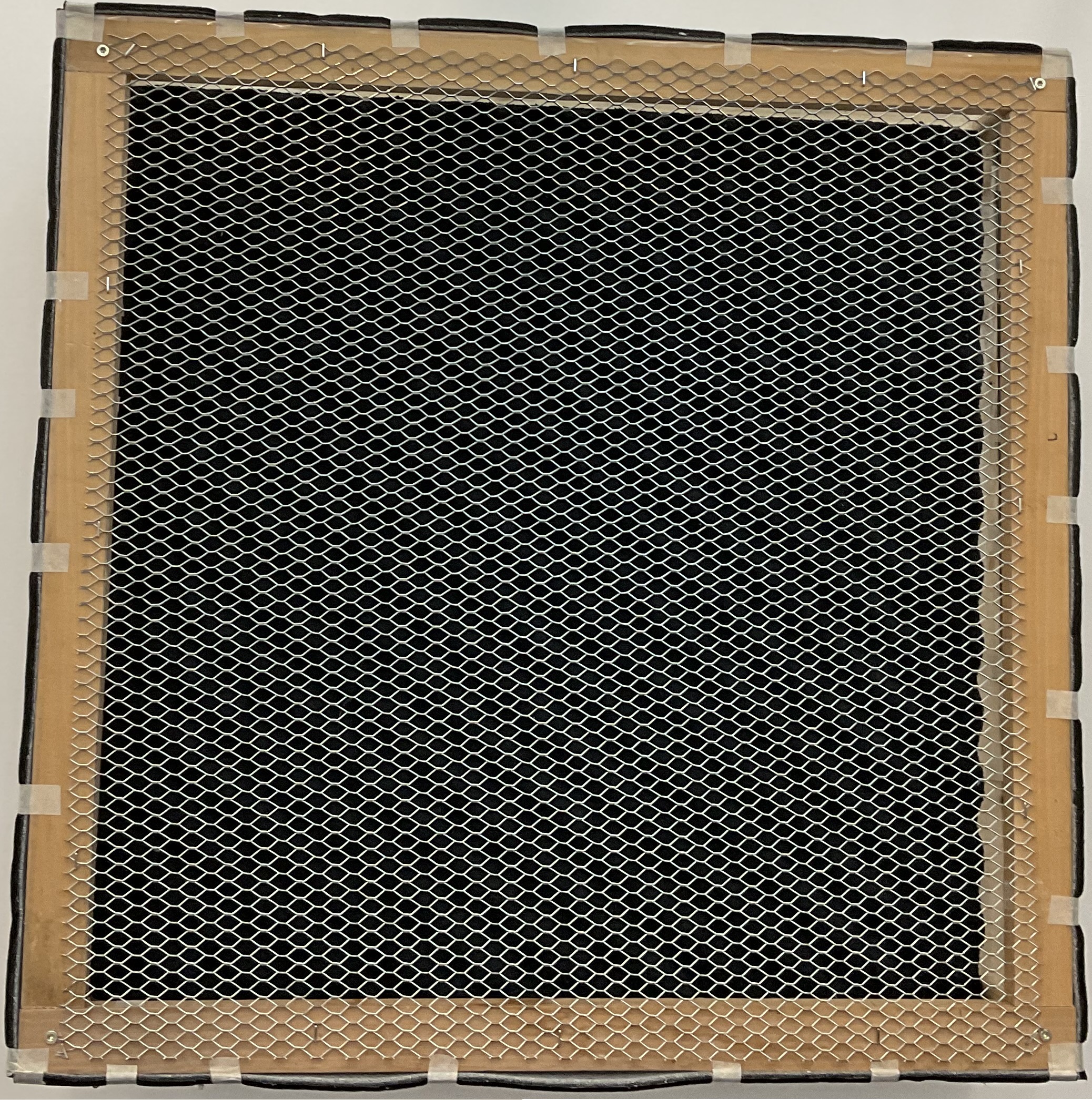}%
}
\caption{Photographs of the materials under test (MUTs), selected to span a wide range of surface roughness and electromagnetic properties. From \cite{Sloane2025XPD}.}
\label{fig:materials_grid}
\vspace{-0.5cm}
\end{figure*}

The baseline XPD model introduced in Section~\ref{subsec:angle-independent} provides an accurate description of the average depolarization behavior in multipath-rich environments. In that formulation, the power-split factor, or equivalently the \gls{xpd}, is assumed to be constant for a given object and therefore independent of the incidence and observation angles. However, this assumption does not generally hold for material-specific depolarization, as also observed in the literature \cite{5062505,10705063,8581035,8870411}.
In this section, we propose a novel analytical model to describe depolarization within the effective roughness framework by explicitly accounting for its dependence on the scattering geometry.
The main idea is to introduce polarization-dependent directivity of the diffuse scattering power, i.e., different exponential factors $\alpha_R$ for the \gls{cp} and \gls{xp} components.

Since the \gls{xpd} is defined at the target level, the total co-polarized and cross-polarized scattered powers must first be derived. This is achieved by summing the contributions from all surface tiles under the ER framework.

As illustrated in Fig.\,\ref{fig:testbed}, the \gls{mut} is discretized into tiles, and the \gls{ds} contribution of the $n$-th tile is evaluated as:
\begin{equation}
\label{eq:Es_n_polarimetric}
\mathbf{E}_{\scriptscriptstyle \mathrm{S}}^{(n)} = |E_{\scriptscriptstyle \mathrm{S,pp}}^{(n)}|e^{-j\beta_{pp}^{(n)}}\; \hat{\mathbf{e}}_{\scriptscriptstyle \mathrm{p}} + |E_{\scriptscriptstyle \mathrm{S,pq}}^{(n)}| e^{-j\beta_{pq}^{(n)}}\; \hat{\mathbf{e}}_{\scriptscriptstyle \mathrm{q}}
\end{equation}
where the intensities of the \gls{cp} and \gls{xp} field components are computed through
\begin{table}[!t]
\caption{Measurement setup parameters (see Fig.\,\ref{fig:testbed}).\label{tab:geom_params}}
\centering
\begin{tabular}{|c||l|c|}
\hline
\textbf{Symbol} & \textbf{Description} & \textbf{Value (m)} \\
\hline
$h_{\scriptscriptstyle \text{RX}}$        & Receiver height            & $1.53$ \\
\hline
$h_{\scriptscriptstyle \text{TX}}$        & Transmitter height         & $1.54$ \\
\hline
$h_{\scriptscriptstyle \text{MUT}}$       & MUT height        & $1.53$ \\
\hline
$L_{\scriptscriptstyle \text{RX}}$        & RX--MUT distance           & $1.50$ \\
\hline
$L_{\scriptscriptstyle \text{TX}}$        & TX--MUT distance           & $0.75$ \\
\hline
$d$ & Distance offset          & $0.07$ \\
\hline
\end{tabular}
\end{table}
\begin{equation}\label{eq:e_s_n}
\left\{
\begin{aligned}
\left|E_{\mathrm{S,pp}}^{(n)}\right|^{2}
= S_{pp}^2 \,A_{p}^{(n)} \,
\frac{f_{pp}\left(\hat{\mathbf{k}}_{s}^{(n)} , \hat{\mathbf{k}}_{i}^{(n)}, \alpha_{\scriptscriptstyle R, pp}\right)}{F_{pp}\left(\hat{\mathbf{k}}_{i}^{(n)}, \alpha_{\scriptscriptstyle R, pp}\right)}\, \mathrm{dS}^{(n)} \\
\left|E_{\mathrm{S,pq}}^{(n)}\right|^{2}
= S_{pq}^2 \,A_{p}^{(n)} \,
\frac{f_{pq}\left(\hat{\mathbf{k}}_{s}^{(n)} , \hat{\mathbf{k}}_{i}^{(n)}, \alpha_{\scriptscriptstyle R, pq}\right)}{F_{pq}\left(\hat{\mathbf{k}}_{i}^{(n)}, \alpha_{\scriptscriptstyle R, pq}\right)}\, \mathrm{dS}^{(n)}
\end{aligned}
\right.
\end{equation}

In \eqref{eq:e_s_n}, $A_{\scriptscriptstyle \text{p}}^{(n)}$ accounts for contributions depending on \gls{tx} power and antenna gain, incidence direction, power reflectivity and \gls{tx}/\gls{rx} distances from the $n$-th tile with the same meaning as in \eqref{eq:e_s}, while
\begin{equation}\label{eq:s_pp_pq}
S_{pp}^2=S_p^2 \cdot (1-\kappa_p^{\scriptscriptstyle \text{AD}}),
\qquad
S_{pq}^2=S_p^2 \cdot \kappa_p^{\scriptscriptstyle \text{AD}}
\end{equation}
are the scattering coefficients for the \gls{cp} and \gls{xp} components, respectively, expressed as a function of the \textit{total scattering coefficient} $S_{p}$ and the \textit{depolarization factor} denoted by $\kappa_p^{\scriptscriptstyle \text{AD}}$, where AD denotes the angle-dependent formulation.

Under the assumption of incoherent scattering from an illuminated object composed of $N$ facets (see Fig.\,\ref{fig:testbed}), that is, with individual contributions having random phases uniformly distributed in $[-\pi,\pi]$, the total \gls{ds} power for the \gls{cp} and \gls{xp} components is given by the superposition of the power contributions scattered by the individual facets, weighted by the factor $K_{\scriptscriptstyle \text{RX}}^{(n)}$ that takes into account the \gls{rx} antenna gain and mismatch losses, i.e.
\begin{equation}\label{eq:e_cp_xp}
\left\{
\begin{aligned}
P_{\scriptscriptstyle S,\text{CP}} \propto \sum_{n=1}^{N}\left(K_{\scriptscriptstyle \text{RX}}^{(n)} \cdot \left|E_{\scriptscriptstyle \mathrm{S,pp}}^{(n)}\right|\right)^{2} \\
P_{\scriptscriptstyle S,\text{XP}} \propto \sum_{n=1}^{N}\left(K_{\scriptscriptstyle \text{RX}}^{(n)}  \cdot \left|E_{\scriptscriptstyle \mathrm{S,pq}}^{(n)}\right|\right)^{2}
\end{aligned}
\right.
\end{equation}
Substituting \eqref{eq:e_s_n}--\eqref{eq:e_cp_xp} into the definition of \gls{xpd} in \eqref{eq:xpd_vitucci}, the proposed angle-dependent \gls{xpd} is obtained as
\begin{equation}\label{eq:xpd_mymodel}
\begin{aligned}
\mathrm{XPD}_p
&=
\frac{1-\kappa_p^{\scriptscriptstyle \text{AD}}}{\kappa_p^{\scriptscriptstyle \text{AD}}}
\cdot
\frac{
\sum\limits_{n=1}^{N}
G_{p}^{(n)}\;
\dfrac{f_{pp}\left(\hat{\mathbf{k}}_{s}^{(n)},\hat{\mathbf{k}}_{i}^{(n)}, \alpha_{\scriptscriptstyle R, pp}\right)}{F_{pp}\left(\hat{\mathbf{k}}_{i}^{(n)}, \alpha_{\scriptscriptstyle R, pp}\right)}\,\mathrm{dS^{(n)}}
}{
\sum\limits_{n=1}^{N}
G_{p}^{(n)}\;
\dfrac{f_{pq}\left(\hat{\mathbf{k}}_{s}^{(n)},\hat{\mathbf{k}}_{i}^{(n)}, \alpha_{\scriptscriptstyle R, pq}\right)}{F_{pq}\left(\hat{\mathbf{k}}_{i}^{(n)}, \alpha_{\scriptscriptstyle R, pq}\right)}\,\mathrm{dS^{(n)}}
}
\end{aligned}
\end{equation}
with
\begin{equation*}
G_{p}^{(n)}=A_{p}^{(n)} \cdot \left(K_{\scriptscriptstyle \text{RX}}^{(n)}\right)^2=\left(\dfrac{ K_{\scriptscriptstyle \text{TX}}^{(n)} K_{\scriptscriptstyle \text{RX}}^{(n)} \Gamma_{p}^{(n)}}{r_{i}^{(n)} r_{s}^{(n)}}\right)^{2}
\cos\left(\vartheta_{i}^{(n)}\right)
\end{equation*}

It can be observed that \eqref{eq:xpd_mymodel} depends on $\kappa_p^{\scriptscriptstyle \text{AD}}$ only as a scalar factor, while the two summations differ solely through the scattering patterns of the \gls{cp} and \gls{xp} components. When the two patterns are identical, \eqref{eq:xpd_mymodel} reduces to \eqref{eq:xpd_vitucci}. Moreover, the proposed formulation is inherently general and can be applied to an arbitrary directive scattering model. It is worth emphasizing that the tiling procedure, i.e., the subdivision of a surface into smaller tiles or facets for \gls{ds} power evaluation, can be applied to any planar surface according to the \gls{er} approach.

\section{Channel Measurement\label{sec:exp_val}}

This section establishes the measurement framework and preprocessing pipeline for validation and polarimetric extension of the effective roughness models. In particular, it describes how super-resolved \glspl{mpc} are extracted from the measurements, allowing the separation of specular and diffuse components. This latter capability is central to our work, as it supports independent calibration of the two contributions, thereby preventing bias in the estimation of diffuse scattered power.

\subsection{Measurement Campaign}\label{sec:meas_camp}

A fully polarimetric bistatic mmWave measurement campaign was conducted at NIST on ten \gls{mut} material samples.

The measurement setup is illustrated in Fig.\,\ref{fig:testbed} and\,\ref{fig:setup_photo}, with all geometric parameters summarized in Tab.\,\ref{tab:geom_params}, and the complete set of \gls{mut}s depicted in Fig.\,\ref{fig:materials_grid}. During the campaign, the \gls{tx} and \gls{rx} remained fixed, whereas the \gls{mut} was mounted on a two-axis positioner that swept azimuth angles from $-15^\circ$ to $75^\circ$ and elevation angles from $0^\circ$ to $75^\circ$, for a total of 266 orientations. This measurement strategy provides a substantially richer angular characterization than conventional azimuth-only rotation campaigns \cite{https://doi.org/10.1155/2020/1583854,8758806,7400901,10137406}. The \gls{mut} is centered at $\mathbf{P}_{\scriptscriptstyle \text{MUT}}$ with surface normal $\hat{\mathbf{n}}_{\scriptscriptstyle \text{MUT}}$, while the rotator reference point $\mathbf{P}_{\text{T/R}}$ is displaced by a distance $d$.

Measurements were acquired in a semi-anechoic environment using a context-aware channel sounder operating at 28.5\,GHz  \cite{Sloane2025XPD}. The \gls{tx} employs a dual-polarized omnidirectional radiation pattern to illuminate the \gls{mut}, while the \gls{rx} consists of a 256-element dual-polarized phased-array antenna for angle-of-arrival (AoA) estimation. All polarization combinations (HH, HV, VH, and VV) are measured at each \gls{mut} orientation. The lidar and camera systems of the channel sounder are used to generate a digital twin (DT) of the \gls{mut}, for subsequent DT-assisted processing.

Super-resolved MPC extraction is performed from the measurement data to obtain delay- and AoA-resolved propagation paths together with full polarimetric characterization. For each \gls{mut} orientation, the delay-resolved channel response of all 256 array elements is sequentially acquired by switching the antennas one at a time. This response is obtained by transmitting a pseudorandom noise (PN) sequence with 2\,GHz chip rate followed by matched filtering at the RX, to generate the delay-resolved response at each antenna. The resulting data are digitally beamformed in postprocessing to generate complex delay--azimuth--elevation profiles for all polarization combinations. The use of switched beamforming~\cite{caudill2021real}, whereby the response of each antenna element is acquired individually prior to digital beamforming, is critical for achieving the required angular resolution and estimation accuracy.

The \gls{sage} algorithm~\cite{papazian2016calibration} is then applied \emph{jointly} to these profiles to extract discrete \gls{mpc}s with common delay and \gls{aoa}, and polarization-dependent complex amplitudes. For each $k$-th orientation, the extracted \gls{mpc} set is
\[
\mathcal{M}(\Omega_k)=\left\{\left(\boldsymbol{a}_j,\tau_j,\mathbf{AoA}_j\right)\right\}_{j=1}^{N_k},
\]
where $\boldsymbol{a}_j =
\left[a_{j,\text{HH}},a_{j,\text{HV}},a_{j,\text{VH}},a_{j,\text{VV}}\right]^\intercal$
is the complex path-amplitude vector over all available polarization pairs,
$\tau_j$ is the propagation delay, and
$\mathbf{AoA}_j =
\left[\theta_j^{\scriptscriptstyle A},\theta_j^{\scriptscriptstyle E}\right]^\intercal$
is the \gls{aoa} vector of the $j$-th MPC. Here, $\theta_j^{\scriptscriptstyle A}$ and $\theta_j^{\scriptscriptstyle E}$ are the azimuth and elevation \gls{aoa}s, respectively, defined with respect to the RX, and $N_k$ denotes the number of extracted \gls{mpc}s for the $k$-th \gls{mut} orientation $\Omega_k$.
The phase information in $a_{j,pq}$ is preserved across all measurements through phase-coherent acquisition enabled by optical synchronization between the \gls{tx} and \gls{rx}. Phase coherence is essential for the complex summation of \gls{mpc} contributions when computing the total diffuse scattered power.

\subsection{Separation of Specular and Diffuse MPCs}
\label{sec:sr_ds_separation}

The method to separate the measured \glspl{mpc} into specular and diffuse components is based on comparing two geometric representations of each propagation event: the incident point associated with each measured \gls{mpc}, and the reference specular point determined from the geometry of the \gls{mut}. The overall procedure consists of three steps: mapping each \gls{mpc} to an incident point on the \gls{mut}, determining the specular point from the MUT geometry, and applying the proposed separation method.

\subsubsection{MPC-to-Incident-Point Mapping}

Each measured \gls{mpc} is mapped to an incident point on the \gls{mut} using its delay and \gls{aoa}, together with the known \gls{tx} and \gls{rx} positions. The delay defines a bistatic constant-delay ellipsoid, while the \gls{aoa} defines a ray originating at the receiver. The intersection of this ray with the ellipsoid and the \gls{mut} surface yields the incident point $\mathbf{p}_j$ associated with the $j$-th \gls{mpc}~\cite{10999612}. This mapping transforms the parametric \gls{mpc} estimates into spatially distributed incident points on the \gls{mut}, as illustrated in Fig.~\ref{fig:Frames}, where each point corresponds to a measured \gls{mpc} colored according to its gain.

\subsubsection{DT-Assisted Processing}
\begin{algorithm}[!ht]
\caption{\gls{sp} and \gls{ds} classification procedure over all \gls{mut} orientations, where $\boldsymbol{a}_j$ and $\mathbf{AoA}_j$ are the complex path gain vector and \gls{aoa} of the $j$-th \gls{mpc} from the $k$-th \gls{mut} orientation.}
\label{alg:sp_cluster_assignment_all}
\begin{algorithmic}[1]
\REQUIRE Set of MUT orientations $\mathcal{I}_{\Omega} = \{\Omega_k\}_{k=1}^{N_\Omega}$, radius $\rho$, path gain threshold $PG_{\mathrm{TH}}$, MPC sets $\{\mathcal{M}(\Omega_k)\}_{k=1}^{N_\Omega}$ with $\mathcal{M}(\Omega_k)=\{\left(\boldsymbol{a}_j,\tau_j,\mathbf{AoA}_j\right)\}_{j=1}^{N_k}$
\ENSURE For each $\Omega_k$: \gls{sp}-MPCs set $\mathcal{S}(\Omega_k)$ and DS-MPCs set $\mathcal{D}(\Omega_k)$

\FOR{$k = 1$ \TO $N_\Omega$}
    \STATE $\mathcal{S}(\Omega_k) \gets$ $\emptyset$
    \STATE $\mathcal{D}(\Omega_k) \gets \mathcal{M}(\Omega_k)$ \COMMENT{Initialize: all MPCs are DS}
    \STATE \COMMENT{Compute theoretical \gls{sp} position $\mathbf{p}_{\mathrm{SP}}$}
    \IF{$\mathbf{p}_{\mathrm{SP}} \notin \mathrm{MUT}$}
        \STATE \COMMENT{No geometric specular point on the MUT $\Rightarrow$ all MPCs are classified DS}
        \STATE \textbf{continue}
    \ENDIF
    \STATE \COMMENT{$\mathbf{p}_j$: projection of each MPC on the MUT}
    \STATE $\mathcal{C} \gets \left\{ j \in \{1,\dots,N_k\} : \left\|\mathbf{p}_j-\mathbf{p}_{\mathrm{SP}}\right\|\le \rho \right\}$
    \IF{$\mathcal{C} = \emptyset$}
        \STATE \COMMENT{No MPC close to the theoretical \gls{sp} $\Rightarrow$ all MPCs are classified DS}
        \STATE \textbf{continue}
    \ENDIF

    \STATE $PG_{j, pp} \gets 10\log_{10}\!\left(|a_{j,pp}|^2\right)$
    \STATE $PG_{\max} \gets \max_{j \in \mathcal{C}} PG_{j, pp}$
    \STATE $\mathcal{S}' \gets \left\{ j \in \mathcal{C} : PG_{\max} - PG_{j, pp} < PG_{\mathrm{TH}} \right\}$
    \IF{$\mathcal{S}' = \emptyset$}
        \STATE \COMMENT{No MPC satisfies the path gain test $\Rightarrow$ all MPCs are classified DS}
        \STATE \textbf{continue}
    \ENDIF

    \STATE $\mathcal{S}(\Omega_k) \gets \mathcal{S}'$ \COMMENT{Chosen \gls{sp} MPCs}
    \STATE $\mathcal{D}(\Omega_k) \gets \mathcal{D}(\Omega_k)\setminus\mathcal{S}'$ \COMMENT{All others are DS MPCs}
\ENDFOR

\end{algorithmic}
\end{algorithm}
The DT of the MUT provides its center, size, and orientation per acquisition. This geometry is used for two purposes: (i) to perform 3D spatial gating of the extracted \glspl{mpc}, ensuring that only propagation paths originating from the \gls{mut} are retained. This rejects contributions from the surrounding environment, edge diffraction, and partial illumination effects that cannot be reliably separated using delay or angle information alone; (ii) to determine the specular reflection point $\mathbf{p}_{\mathrm{SR}}$ based on bistatic specular reflection geometry, using the known \gls{tx} and \gls{rx} positions. When a valid specular solution exists within the bounds of the \gls{mut}, it serves as the reference location for the specular component.

\subsubsection{Separation Method}
\begin{figure*}[!t]
\centering
\includesvg[width=7in]{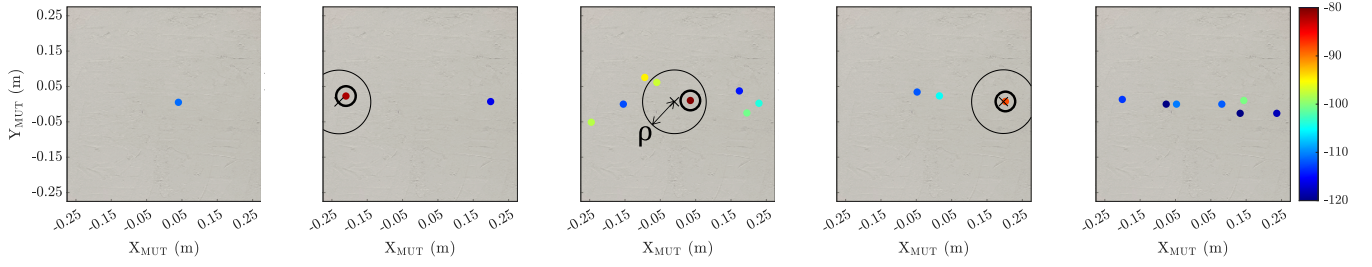}
\caption{Example of \gls{sp} and \gls{ds} classification using Algorithm\,\ref{alg:sp_cluster_assignment_all} for the mortar and VV dataset. Here, \gls{mut} rotates from 10 to 50 degrees in 10-degree steps, with tilt fixed at zero. The large circles refer to the region where specular reflection is expected to be present, while the circled \gls{mpc}s are the ones classified as the specular reflection.\label{fig:Frames}}
\vspace{-0.3cm}
\end{figure*}
The proposed separation method compares the measured incident points $\mathbf{p}_j$ with the DT-based specular point $\mathbf{p}_{\mathrm{SP}}$. For each \gls{mut} orientation, candidate specular \gls{mpc}s are identified as those whose incident points lie within a radius $\rho$ of $\mathbf{p}_{\mathrm{SP}}$.
An example of a specular cluster within the \gls{mut} and a potential specular candidate $\mathbf{p}_j$ are shown in Fig. \ref{fig:testbed}.
Among these candidates, only the paths whose gain value is not below the threshold quantity $PG_{\mathrm{TH}}$ compared to the strongest candidate are retained as specular components, while all remaining \gls{mpc}s are assigned to the diffuse contribution.
This joint spatial and power-based criterion ensures that the specular cluster remains physically meaningful and avoids overestimation due to nearby diffuse paths. The complete classification procedure is summarized in Algorithm\,\ref{alg:sp_cluster_assignment_all}.

The thresholds $\rho$ and $PG_{\mathrm{TH}}$ are selected conservatively. The radius $\rho$ is chosen to be on the order of the projected \gls{ffz}. For orientations where specular reflection is expected (rotation angles between $20^\circ$ and $40^\circ$ and tilt angles between $0^\circ$ and $15^\circ$), the projected \gls{ffz} is nearly circular, with radius between 7.3 and 8.2\,cm. In practice, finite angular resolution and surface roughness enlarge the effective specular region. Accordingly, $\rho$ is adapted to the material, ranging from 7\,cm for smooth surfaces (e.g., plexiglass) up to 15\,cm for rough surfaces (e.g., brick).
Due to the complex geometry of the fence structures, relatively high values of $\rho$ were assigned to account for the high spatial spreading of the scattered power.
A circular region is used in place of the exact \gls{ffz} ellipse for simplicity while preserving physical consistency.
The threshold $PG_{\mathrm{TH}}$ reflects the expected separation between specular and diffuse contributions. Tighter thresholds are used for rough materials, where the two contributions are closer in magnitude, while smoother materials allow for a less restrictive threshold. The resulting values of $\rho$ and $PG_{\mathrm{TH}}$ are summarized in Table~\ref{tab:clustering_thresholds}.
\begin{table}[!t]
\centering
\caption{Clustering thresholds per material.}
\label{tab:clustering_thresholds}
\renewcommand{\arraystretch}{1.1}
\setlength{\tabcolsep}{6pt}
\begin{tabular}{|l||c|c|}
\hline
\textbf{Material} & \textbf{$\rho$ (cm)} & \textbf{$PG_{TH}$ (dB)} \\
\hline
Brick                  & 15 & 5 \\
Shingles               & 10 & 5 \\
Tile                   & 10 & 5 \\
Carpet                 & 10 & 5 \\
Drywall                & 7 & 10 \\
Mortar                 & 7 & 10 \\
Plexiglass             & 7 & 15 \\
Plywood                & 15 & 5 \\
Fence 45$^\circ$       & 25 & 15 \\
Fence 0$^\circ$        & 25 & 15 \\
\hline
\end{tabular}
\end{table}
An example of the resulting classification is shown in Fig.~\ref{fig:Frames} for mortar in VV polarization across multiple \gls{mut} orientations. The local coordinate system $(X_{\mathrm{MUT}}, Y_{\mathrm{MUT}})$ is defined on the \gls{mut} surface, and the circular region centered at $\mathbf{p}_{\mathrm{SP}}$ identifies the candidate specular cluster. The dominant path within this region is classified as specular, while the remaining paths are assigned to the diffuse component.

\section{Model Calibration and Validation}\label{sec:calibration}

This section describes the parameter calibration and validation of the proposed model. The first subsection presents the calibration procedure, while the second analyzes the estimated parameters and resulting model performance.

\begin{figure}[!t]
\centering
\includesvg[width=3.5in]{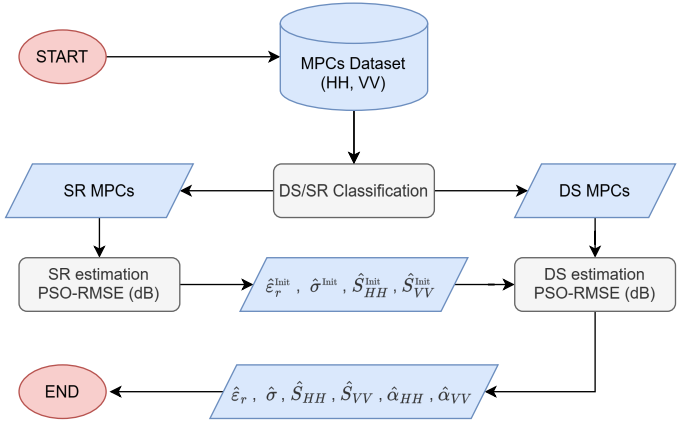}
\caption{Flowchart of the two-stage calibration procedure for the \gls{cp} component.}
\label{fig:FlowchartCPcalib}
\vspace{-2mm}
\end{figure}

\subsection{Calibration Procedure}

The model parameters for calibration are the relative permittivity, conductivity, scattering coefficients, exponential factors, and depolarization factors of the MUT. The calibration is performed sequentially, with the dominant power-related parameters estimated from the \gls{cp} data, followed by the depolarization-related quantities inferred from the \gls{xp} response.
The workflow of the calibration procedure is summarized in Fig.\,\ref{fig:FlowchartCPcalib}.

Regarding power calibration, the raw measured path gain of the \gls{ds} component for the $k$-th \gls{mut} orientation $\Omega_k$ and $pq$ polarization is derived by coherently summing the complex amplitude of each \gls{mpc} $a_{i,pq}(\Omega_k)$, such as
\[
PG_{pq}(k)
=
\left|
\sum_{i \in \mathcal{D}(\Omega_k)}
a_{i,pq}(\Omega_k)
\right|^2 \,,
\]
where $p,q \in \{\text{H}, \text{V}\}$.
Similarly, path gain for the \gls{sp} component is obtained by coherently summing the specular \glspl{mpc}, i.e. by summing $a_{i,pq}(\Omega_k)$ with $i \in \mathcal{S}(\Omega_k)$.

Subsequently, the measured \gls{xpd}, which we recall represents the ratio between \gls{cp} ($PG_{pp}$) and \gls{xp} ($PG_{pq}$) components, is obtained as
\[
\text{XPD}_p(k) = \frac{PG_{pp}(k)}{PG_{pq}(k)} \, .
\]
\begin{table}[!t]
\centering
\caption{PSO algorithm parameters.}
\label{tab:PSOparams}
\renewcommand{\arraystretch}{1.1}
\setlength{\tabcolsep}{6pt}
\begin{tabular}{|l||c|}
\hline
\textbf{Parameter} & \textbf{Value} \\
\hline
Inertia coefficient    & 0.7 \\
Cognitive coefficient  & 1.5 \\
Social coefficient     & 1.5 \\
Swarm size             & 30  \\
Iterations             & 20  \\
\hline
\end{tabular}
\end{table}
Since the effective roughness models describe the large-scale angular behavior of the diffuse power, direct calibration against raw measurements is inappropriate, as the measured response contains small-scale fluctuations arising from the coherent combination of individual \glspl{mpc} with randomly varying phases \cite{parsons_mobile_2001}. As such, the measured data are first processed to extract the \gls{lst} of diffuse power. The preprocessing procedure is described in detail in the Appendix.

\begin{table*}[!ht]
\caption{Comprehensive set of material-specific estimated parameters for the G-\gls{rer} model coupled with the angle-dependent \gls{xpd} model.}
\label{tab:material_params_cp_xp_metrics3}
\centering
\footnotesize
\setlength{\tabcolsep}{2.5pt}
\renewcommand{\arraystretch}{1.1}

\rowcolors{1}{rowgray}{white}

\begingroup
\arrayrulecolor{black}

\resizebox{\textwidth}{!}{%
\begin{tabular}{|l||c|c||c|c|c|c||c|c|c||c|c|c|c||c|c|c|}
\hline
\hiderowcolors
\textbf{Material} &
$\mathbf{\varepsilon_r}$ &
$\mathbf{\sigma}$ &
\multicolumn{7}{c||}{\textbf{H}} &
\multicolumn{7}{c|}{\textbf{V}} \\
\noalign{\setlength{\arrayrulewidth}{0.8pt}}
\cline{4-10}\cline{11-17}
\noalign{\setlength{\arrayrulewidth}{0.4pt}}
 & & &
\multicolumn{4}{c||}{\textbf{DS parameters}} &
\multicolumn{3}{c||}{\textbf{Performance metrics}} &
\multicolumn{4}{c||}{\textbf{DS parameters}} &
\multicolumn{3}{c|}{\textbf{Performance metrics}} \\
\noalign{\setlength{\arrayrulewidth}{0.8pt}}
\cline{4-10}\cline{11-17}
\noalign{\setlength{\arrayrulewidth}{0.4pt}}
 & & &
\textbf{$\mathbf{\alpha_R}$, HH} &
\textbf{$\mathbf{\alpha_R}$, HV} &
\textbf{$\mathbf{S_H}$} &
\textbf{$\mathbf{\kappa_H^{\scriptscriptstyle \text{AD}}}$} &
\textbf{RMSE$\mathbf{_{HH}}$ (dB)} &
\textbf{RMSE$\mathbf{_{HV}}$ (dB)} &
\textbf{SS$\mathbf{_H}$} &
\textbf{$\mathbf{\alpha_R}$, VV} &
\textbf{$\mathbf{\alpha_R}$, VH} &
\textbf{$\mathbf{S_V}$} &
\textbf{$\mathbf{\kappa_V^{\scriptscriptstyle \text{AD}}}$} &
\textbf{RMSE$\mathbf{_{VV}}$ (dB)} &
\textbf{RMSE$\mathbf{_{VH}}$ (dB)} &
\textbf{SS$\mathbf{_V}$} \\
\hline
\showrowcolors

Brick                  & 3.219 & 0.050  & 5.667 & 2.503 & 0.40 & 0.0131 & 4.55 & 4.63 & 0.56 & 6.467 & 1.180 & 0.36 & 0.0311 & 3.04 & 3.58 & 0.68 \\
Shingles               & 3.718 & 0.097  & 5.262 & 1.105 & 0.35 & 0.0199 & 4.35 & 4.17 & 0.91 & 5.152 & 1.496 & 0.40 & 0.0237 & 4.18 & 4.41 & 0.59 \\
Tile                   & 3.806 & 0.167  & 4.738 & 2.021 & 0.55 & 0.0170 & 4.42 & 4.34 & 0.81 & 5.201 & 1.221 & 0.46 & 0.0238 & 3.74 & 3.81 & 0.82 \\
Carpet                 & 2.180 & 0.095  & 2.478 & 1.724 & 0.25 & 0.0058 & 4.41 & 4.79 & 0.00 & 4.255 & 2.363 & 0.25 & 0.0110 & 3.99 & 4.59 & 0.64 \\
Drywall                & 3.315 & 0.076  & 3.507 & 2.098 & 0.30 & 0.0104 & 3.28 & 3.64 & 0.57 & 1.874 & 1.806 & 0.25 & 0.0227 & 3.06 & 3.65 & 0.00 \\
Mortar                 & 3.040 & 0.135  & 3.945 & 0.000 & 0.35 & 0.0133 & 4.04 & 4.37 & 0.80 & 1.134 & 0.455 & 0.30 & 0.0265 & 5.10 & 4.83 & 0.36 \\
Plexiglass             & 4.000 & 0.439  & 3.830 & 3.105 & 0.17 & 0.0115 & 4.48 & 4.72 & 0.27 & 1.726 & 1.398 & 0.15 & 0.0119 & 3.87 & 3.58 & 0.00 \\
Plywood                & 2.563 & 0.032  & 3.768 & 1.056 & 0.30 & 0.0081 & 3.02 & 4.17 & 0.53 & 1.545 & 2.318 & 0.20 & 0.0043 & 3.79 & 4.50 & 0.14 \\
Fence 45$^\circ$       & 1.611 & 24.00  & 0.152 & 1.525 & 0.17 & 0.2486 & 5.23 & 7.75 & 0.75 & 0.180 & 0.000 & 0.17 & 0.1838 & 5.58 & 6.39 & 0.11 \\
Fence 0$^\circ$        & 11.06 & 1000.0 & 0.000 & 0.000 & 0.10 & 0.0401 & 8.17 & 9.01 & 0.00 & 0.000 & 0.000 & 0.12 & 0.3319 & 6.96 & 8.10 & 0.00 \\
\hline
\end{tabular}%
}
\endgroup
\end{table*}

The model parameters are then calibrated by jointly estimating the specular and diffuse contributions. For the specular component, $\varepsilon_r$ and $\sigma$ are estimated using a Fresnel-based formulation similar to \cite{1433205}.

For the diffuse component, the \gls{cp} scattering coefficient $S_{pp}$ is estimated by matching the measured \gls{cp} diffuse component with the first equation in \eqref{eq:e_s_n}, which corresponds to assigning a depolarization factor $\kappa=0$ in the effective roughness polarimetric model. Similarly, the \gls{xp} scattering coefficient $S_{pq}$ is obtained using the second equation in \eqref{eq:e_s_n}, which means assigning a depolarization factor $\kappa=1$ in the model.

Using \eqref{eq:s_pp_pq}, the estimated depolarization factor $\kappa_p^{\scriptscriptstyle \text{AD}}$ and overall scattering coefficient $S_p$ are then obtained as
\begin{equation}
    \kappa_p^{\scriptscriptstyle \text{AD}} = \frac{S_{pq}^2}{S_{pq}^2 + S_{pp}^2}, \quad
    S_p = \sqrt{S_{pq}^2 + S_{pp}^2}.
\end{equation}
The parameter estimation is carried out using \gls{pso}. A complete list of the \gls{pso} settings is provided in Tab.~\ref{tab:PSOparams}. A full joint optimization over the combined \gls{sp} and \gls{ds} parameter sets would be computationally prohibitive. Each dataset contains 266 \gls{mut} orientations, and with 30 particles, the calibration requires 159600 \gls{rt} simulations for the \gls{ds} component and 600 for the \gls{sp} component, even for a single polarization. This computational burden justifies the adopted sequential calibration strategy.

The \gls{rmse} is used as the loss function for the path gain estimation, since the objective is to accurately reproduce the trend of the \gls{ds} power component as a function of the \gls{mut} orientation. The \gls{rmse} is computed for each $pq$ polarization component as
\begin{equation}
\label{eq:RMSE}
\text{RMSE}=\sqrt{\frac{1}{N_\Omega}\sum_{k=1}^{N_\Omega}\left(\overline{PG}_{pq}(k)-\widehat{PG}_{pq}(k)\right)^2}
\end{equation}
where $N_\Omega$ is the number of \gls{mut} orientations, $\overline{PG}_{pq}(k)$ is the average measured path gain (\gls{lst}) for the $k$-th orientation, and $\widehat{PG}_{pq}(k)$ is the estimated path gain obtained using the calibrated effective roughness model.

Regarding calibration of \gls{xp} parameters, the goal of the proposed model in \eqref{eq:xpd_mymodel} is to improve upon the baseline \gls{xpd} model in \eqref{eq:xpd_vitucci}. For this reason, the calibration is performed using a skill score metric, which directly quantifies the predictive improvement of the proposed model with respect to the baseline. The skill score is defined as
\begin{equation}
\label{eq:skill_score}
    \mathrm{SS}
    = 1 - \frac{\sum_{\ell=1}^{N_\Omega}\left(\overline{y}_\ell-\hat{y}^{\scriptscriptstyle \text{AD}}_{\ell}\right)^2}
    {\sum_{\ell=1}^{N_\Omega}\left(\overline{y}_\ell-\hat{y}_{\ell}\right)^2},
\end{equation}
where ${\overline{y}}_\ell$ is the measured (\gls{lst}) \gls{xpd}, $\hat{y}^{\scriptscriptstyle \text{AD}}_{\ell}$ is the prediction of the proposed angle-dependent model, and $\hat{y}_{\ell}$ is the prediction of the baseline model. All quantities in \eqref{eq:RMSE}, \eqref{eq:skill_score} are converted to decibels (dB) prior to evaluation.
\begin{figure*}[t]
\centering
\subfloat[HH\label{fig:Brick_HH}]{%
  \includesvg[width=0.48\textwidth]{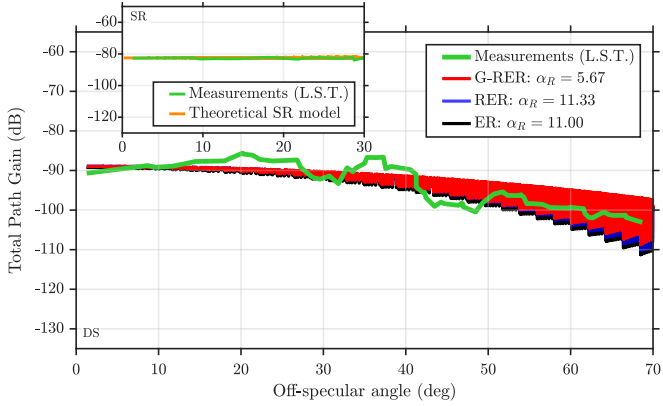}%
}\hfill
\subfloat[HV\label{fig:Brick_HV}]{%
  \includesvg[width=0.48\textwidth]{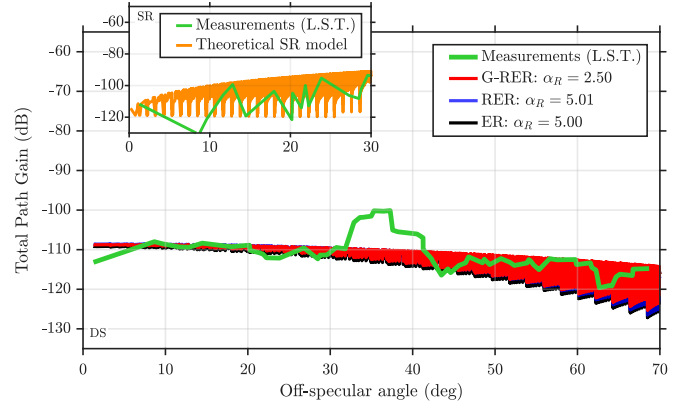}%
}\\[0.5ex]
\subfloat[VV\label{fig:Brick_VV}]{%
  \includesvg[width=0.48\textwidth]{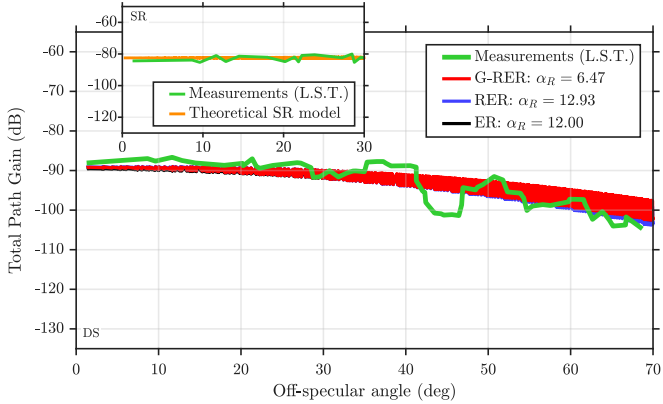}%
}\hfill
\subfloat[VH\label{fig:Brick_VH}]{%
  \includesvg[width=0.48\textwidth]{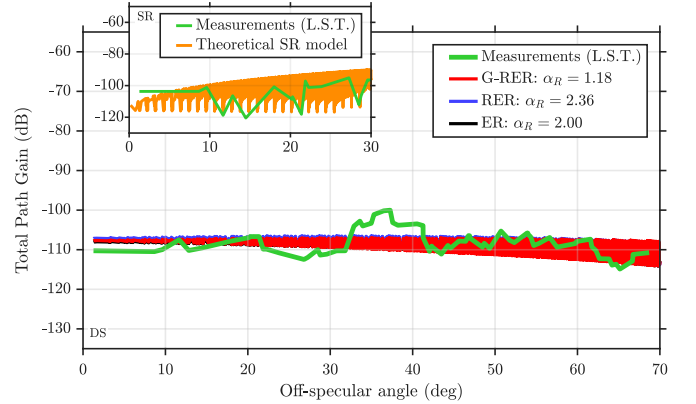}%
}
\caption{Total diffuse power and specular reflection for both \gls{cp} and \gls{xp} components for brick. The green pattern represents the large-scale trend (i.e., L.S.T.) of the total power, while the red, blue, and black patterns represent the predictions using the ER, RER, and G-RER directive patterns. The inset shows the corresponding measured and predicted specular reflection power, where the orange curve represents the theoretical specular reflection model.}
\label{fig:Brick_TOTPG}
\vspace{-0.5cm}
\end{figure*}

\subsection{Parameter Analysis\label{sec:valid}}

\begin{figure*}[t]
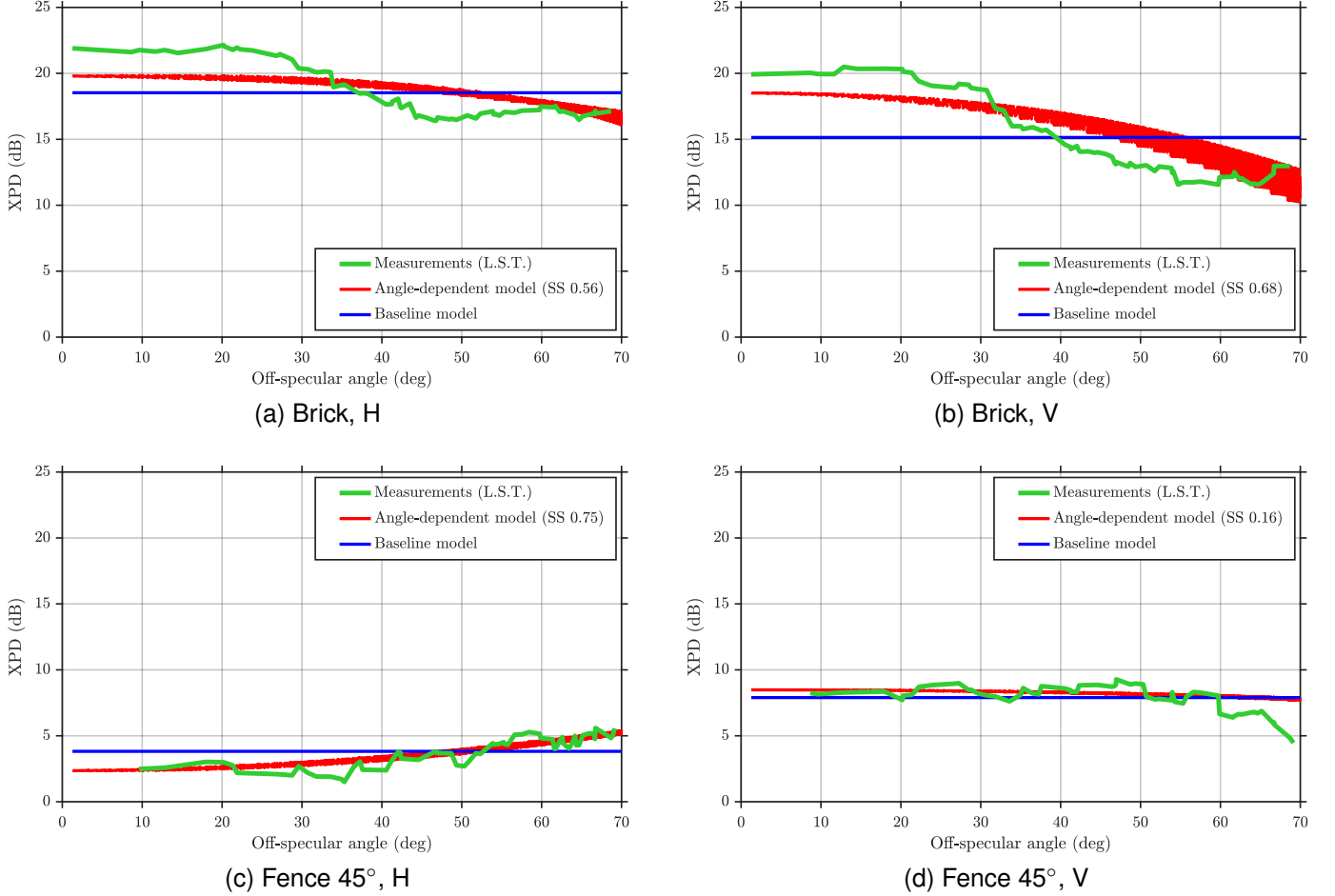

\centering
\subfloat[Brick, H\label{fig:Brick_XPD_H}]{%
  \includesvg[width=0.48\textwidth]{Pictures/SVG/Brick_XPD_H}%
}\hfill
\subfloat[Brick, V\label{fig:Brick_XPD_V}]{%
  \includesvg[width=0.48\textwidth]{Pictures/SVG/Brick_XPD_V}%
}\\[0.5ex]
\subfloat[Fence 45$^\circ$, H\label{fig:Steel_fence_45deg_XPD_H}]{%
  \includesvg[width=0.48\textwidth]{Pictures/SVG/Steel_fence_45deg_XPD_H}%
}
\hfill
\subfloat[Fence 45$^\circ$, V\label{fig:Steel_fence_45deg_XPD_V}]{%
  \includesvg[width=0.48\textwidth]{Pictures/SVG/Steel_fence_45deg_XPD_V}%
}
\caption{Measured and estimated \gls{xpd} for two different materials: brick and fence 45$^\circ$. The results are obtained using the tuned G-\gls{rer} model.}
\label{fig:xpds}
\vspace{-0.5cm}
\end{figure*}

This subsection presents the parameter estimation results obtained from the calibration procedure. Table~\ref{tab:material_params_cp_xp_metrics3} reports the estimated dielectric properties,
relative permittivity and conductivity, together with the scattering and depolarization parameters for all polarization combinations (HH, HV, VV, and VH).

The estimated permittivity and conductivity values are physically consistent and in good overall agreement with the literature across all materials. For brick ($\varepsilon_r = 3.219$), drywall ($\varepsilon_r = 3.315$), and plywood ($\varepsilon_r = 2.563$), the estimates align closely with both ITU-R P.2040-3~\cite{ITU-R-P2040-3} and the comprehensive wideband dataset in~\cite{Zhekov_APM_2020_BuildingMaterialsUWB}, which reports real permittivity values of approximately 3 to 4 for brick, 2 to 3 for plasterboard, and plywood across
0.2 to 67\,GHz. Shingles ($\varepsilon_r = 3.718$) and tile ($\varepsilon_r = 3.806$) fall within the range typically reported for dense construction composites at mmWave
frequencies~\cite{ITU-R-P2040-3}, consistent with their relatively high surface density and heterogeneous composition. For mortar, the estimate ($\varepsilon_r = 3.040$) lies above the range
of 1.91 to 2.23 reported in~\cite{Jawad_PIERL_2020_MortarDielectric} for fully dried laboratory samples measured from 8 to 32\,GHz; this discrepancy is consistent with the known strong
sensitivity of mortar permittivity to mix composition and residual moisture documented therein, and the higher value obtained here likely reflects the denser and thicker construction-grade sample used in
our campaign. For plexiglass, the estimated $\varepsilon_r = 4.0$ exceeds the 2.60 to 2.62 reported for PMMA in the 9.5–75\,GHz range in~\cite{Givot_EuCAP_2021_PolymerPermittivityLoss} and
the $\approx$2.6 reported in~\cite{Zhekov_APM_2020_BuildingMaterialsUWB}. We attribute this increased value to the particular sample used in our campaign, which is likely denser than typical PMMA and
thus behaves as a stronger reflector. In fact, the inferred relative permittivity is closer to that of a glass-like material ($\varepsilon_r = 6.27$, from \cite{ITU-R-P2040-3}). The carpet permittivity ($\varepsilon_r = 2.180$) is physically reasonable and consistent with dense textile substrates surveyed
in~\cite{Yamada_Textiles_2022_DielectricTextileReview}, where dielectric constants of 1.1 to 1.7 are reported for woven and knitted fabrics at microwave frequencies, with higher values expected
for thicker, backed carpets such as the one used here. Finally, the two fence panels exhibit markedly different dielectric characters: the $45^{\circ}$ fence ($\varepsilon_r = 1.611$,
$\sigma = 24.00$\,S/m) reflects the dominant metallic wire contribution of its periodic structure, while the $0^{\circ}$ fence ($\varepsilon_r = 11.06$, $\sigma = 1000$\,S/m) yields a much higher
effective permittivity and conductivity, consistent with the strong resonant EM coupling that occurs when the incident field is aligned with the wire orientation, a well-known behavior of
periodic wire-grid structures.

The scattering parameters, rather, are grouped according to the TX polarization (H and V). Results are reported for the G-\gls{rer} model only, as differences with ER and RER were found to be marginal.

For convenience, we adopt the material grouping in~\cite{Sloane2025XPD}:
\begin{itemize}
\item \textit{Rough} materials: tile, brick, shingles, mortar;
\item \textit{Smooth} materials: plywood, drywall, plexiglass;
\item \textit{Permeable} materials: fence 45$^\circ$, fence 0$^\circ$.
\end{itemize}

Rough materials generally exhibit larger scattering coefficients and exponential factors. The relationship between H and V polarization parameters is material-dependent and the fitted
exponents are not interchangeable across polarizations: for rough materials such as brick, tile, and carpet, $\alpha_{\text{VV}}$ exceeds $\alpha_{\text{HH}}$, while the opposite holds for
smooth materials such as drywall, mortar, and plywood. In most cases, $S_{\text{H}}$ is greater than $S_{\text{V}}$, implying stronger diffuse scattering under H polarization.

With this classification in mind, in the following, we aim to study the trend of \gls{ds} power and \gls{xpd} as functions of the off-specular angle.
Consistent with \cite{Sloane2025XPD}, the off-specular angle $\psi$ is defined with respect to the center of the \gls{mut}. In particular, $\psi$ is computed for each \gls{mut} orientation through \eqref{eq:off-spec}.

As an example, Fig.~\ref{fig:Brick_TOTPG} shows the behaviour of \gls{ds} path gain for brick as a function of the off-specular angle, and for all the \gls{cp} and \gls{xp} polarization combinations. The \gls{lst} of measured \gls{ds} power (green curve) is compared with the \gls{er} models presented in Section II. After calibration, the trend is quite well reproduced, especially by the \gls{g-rer} model (red curve), and with limited values of the exponent $\alpha_R$.

Furthermore, the trend of the specular component (measured vs theoretical) is shown as a reference in the top box, for each of the 4 plots. As expected, the specular \gls{cp} component is very stable and dominant for the angle combinations for which it exists, while the specular \gls{xp} component is highly oscillatory and always comparable to the diffuse component. For angle combinations in which the specular component does not exist, the field depolarization is solely determined by the diffuse component.

More in detail, Fig.~\ref{fig:xpds} shows the trend of \gls{xpd} for the \gls{ds} component, considering both H and V transmit polarizations, and two different material samples, brick and $45^\circ$ fence: the \gls{lst} of measured \gls{xpd} (green curve) is compared here with the calibrated angle-dependent \gls{xpd} model (red curve), whereas the baseline \gls{xpd} model (blue line, constant with the off-specular angle), is also reported as a reference.

Although a clear trend of \gls{ds} path gain and \gls{xpd} with off-specular angle is observed, neither quantity is uniquely determined by it, and this explains why the curves in Fig.s ~\ref{fig:Brick_TOTPG} and \ref{fig:xpds} are not strictly monotonic. This is due to two factors: (i) the scattered field in \eqref{eq:e_s} depends not only on the off-specular angle but also on the incidence elevation angle $\vartheta_i$, and (ii) since the \gls{mut} is not fully in the far field of the \gls{tx}, the local reflectivity $\Gamma_{p,n}^2$ varies across the illuminated surface, introducing dispersion in the observed trends.

Depolarization exhibits a clear material dependence consistent with the scattering behavior. Smooth materials tend to show higher and flatter \gls{xpd} curves, whereas rough materials exhibit lower and more angle-dependent \gls{xpd}, indicating stronger depolarization. For rough materials, $\text{XPD}_{\scriptscriptstyle H}$ typically varies more sharply with off-specular angle, suggesting that the co-polarized component is more directive than the cross-polarized one (see Fig.~\ref{fig:Brick_XPD_H} and Fig.~\ref{fig:Brick_XPD_V}).
In many cases, \gls{xpd} decreases with increasing off-specular angle, indicating increased transfer of co-polarized power into the cross-polarized component. This suggests that the cross-polarized component follows a more random, Lambertian-like behavior. Similar trends have been reported in \cite{5062505, 8581035, 8870411}. In some cases, however, the opposite behavior is observed, where \gls{xpd} increases with off-specular angle. This occurs, for example, for the $45^\circ$ fence in the H case (see Fig.~\ref{fig:Steel_fence_45deg_XPD_H}) and for mortar in the V case, and has also been reported in \cite{10705063}.

The fence materials represent a special case for both diffuse scattering and depolarization. Their $S_{\scriptscriptstyle H}$ and $S_{\scriptscriptstyle V}$ values are significantly smaller than those of the other materials, indicating strong transmission through the apertures and weaker diffuse scattering. At the same time, their depolarization factors are among the largest, showing that periodic, partially transparent structures can strongly depolarize the field. The near interchange of $\kappa_{\scriptscriptstyle H}^{\scriptscriptstyle \text{AD}}$ and $\kappa_{\scriptscriptstyle V}^{\scriptscriptstyle \text{AD}}$ between orientations reflects the rotation of the periodic structure and the resulting change in boundary conditions, enabling more complex \gls{em} coupling. Additionally, for both fence cases, the diffuse power increases with off-specular angle, unlike typical rough-surface scattering. A plausible explanation is that, at small off-specular angles, more incident power is transmitted through the apertures, whereas at larger angles the apertures appear effectively smaller, reducing transmission and redirecting more power into scattering, consistent with power conservation.

Finally, calibration accuracy is quantified in Table~\ref{tab:material_params_cp_xp_metrics3} for all the materials using the RMSE and the skill score defined earlier. The RMSE values for total diffuse power are generally low across all polarization components. The skill score analysis indicates that the proposed \gls{xpd} model consistently improves upon the baseline, yielding positive scores for nearly all materials. The improvement is most significant when \gls{xpd} exhibits a pronounced dependence on the off-specular angle, for which the baseline constant-\gls{xpd} model becomes inadequate. In contrast, when \gls{xpd} is relatively flat, as for plexiglass, the baseline already performs reasonably well, and the achievable improvement is limited.

\section{Conclusion}\label{sec:conclusion}

This paper presented a comprehensive experimental validation of the Effective Roughness (ER) diffuse scattering framework at 28 GHz using a fully polarimetric bistatic measurement campaign conducted at NIST across ten common building materials. A central contribution is the ability to spatially separate specular and diffuse scattering components through super-resolved MPC extraction combined with lidar-assisted digital twin (DT) geometry. This capability enables independent characterization of the two mechanisms, thereby eliminating bias in the estimation of diffuse scattered power that is inherent in aggregate measurements.

The results demonstrate that, when properly parameterized, the ER framework accurately predicts the total diffuse-scattered power across all polarization combinations and materials considered. The extracted dielectric and scattering parameters are physically consistent and align with values reported in the literature, supporting the robustness of the proposed calibration methodology.

Beyond power modeling, this work introduced a novel angle-dependent XPD formulation within the ER framework, which captures the dependence of depolarization on the scattering geometry. In contrast to conventional models that assume a constant depolarization factor, the proposed approach allows distinct angular behavior for co- and cross-polarized components. This extension consistently improves predictive performance over the baseline model, as quantified by positive skill scores across nearly all materials. Naturally, the improvement is most pronounced for materials exhibiting strong angular variation in XPD, while remaining consistent with the baseline in cases where depolarization is approximately geometry-independent.

Overall, the results establish the ER framework as a physically consistent and practically accurate approach for jointly modeling diffuse scattering power and polarization effects at mmWave frequencies. The material-specific parameter sets and modeling methodology developed in this work are directly compatible with ray-tracing tools, providing a pathway toward improved channel modeling fidelity in next-generation wireless system simulations.

\appendix
\label{appendix:A}

\begin{figure*}[!t]
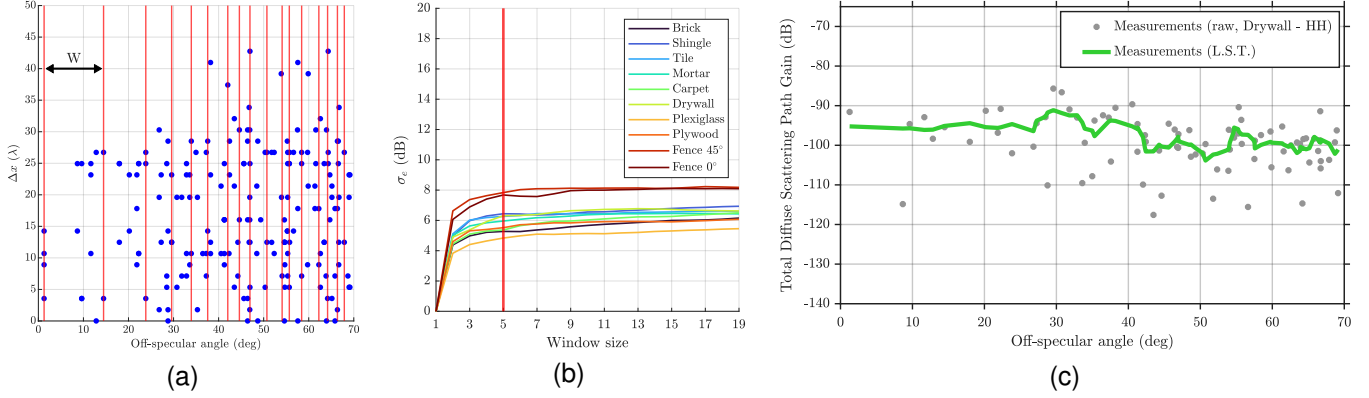

\centering

\makebox[\textwidth][c]{%
\hspace*{-0.04\textwidth}%
\subfloat[]{%
\raisebox{-0.5\height}{%
\includesvg[width=0.26\textwidth]{Pictures/SVG/PathLengthDrywall}}%
\label{fig:PathLengthDrywallH}}
\hspace{0.015\textwidth}
\subfloat[]{%
\raisebox{-0.5\height}{%
\includesvg[width=0.26\textwidth]{Pictures/SVG/LSF_SSF_sigma_HH}}%
\label{fig:LSF_SSF_sigma_HH}}
\hspace{0.015\textwidth}
\subfloat[]{%
\raisebox{-0.5\height}{%
\includesvg[width=0.42\textwidth]{Pictures/SVG/Drywall_HH_Raw_LSF}}%
\label{fig:Carpet_HH_Raw_LSF}}
}

\caption{(a) Path length difference as a function of off-specular angle for drywall; (b) standard deviation of the residual for different window sizes; (c) example of raw and large-scale diffuse power for drywall (HH polarization), illustrating the preprocessing used to extract the large-scale component.}
\label{fig:LSF_SSF_combined}
\vspace{-2mm}
\end{figure*}

\section{Large-Scale Trend Extraction from Diffuse Scattering Power}

This appendix describes the methodology used to extract the \gls{lst} from the measured total \gls{ds} power.
To isolate the \gls{lst} component, we employ a sliding-window averaging estimator \cite{554753,6805659,LYU2024102249}. Among several alternatives (median sliding window, fixed average window, and fixed median window), the moving average was found to provide the most consistent performance across all materials, while avoiding systematic underestimation of the diffuse power.
Importantly, the averaging is performed over orientation indices rather than directly over the off-specular angle. This choice is motivated by the non-stationarity of the \gls{ssf} component with respect to the scattering geometry, as discussed next.

\subsection{Non-Stationarity of the SSF Component}

The \gls{ssf} arises from the coherent superposition of multiple MPCs, whose relative phases depend on the propagation geometry. As the MUT orientation changes, the path length differences between MPCs vary, leading to a non-stationary interference pattern.
To illustrate this, we define the path length difference between two MPCs as
\begin{equation}
    \Delta x_{i,j} = |r_i - r_j|,
\end{equation}
where $r_i$ and $r_j$ are the corresponding path lengths.
Larger path length differences result in more rapid oscillations in the received power.
For each MUT orientation, the path length difference is computed relative to the minimum propagation distance
\begin{equation}
    L_{\text{min}} = \min_{k \, \in \, \mathcal{I}_{\Omega}} \,\left\{ ||\mathbf{P}_{\text{TX}} - \mathbf{P}_{\text{MUT}}(k)|| + ||\mathbf{P}_{\text{RX}} - \mathbf{P}_{\text{MUT}}(k)|| \,\right\},
\end{equation}
where $\mathbf{P}_{\text{TX}}$, $\mathbf{P}_{\text{RX}}$, and $\mathbf{P}_{\text{MUT}}(k)$ denote the TX, RX, and MUT positions, respectively. Then, the path length difference for the $j$-th \gls{mpc} $\Delta x_j$ is
\begin{equation}
    \Delta x_j = r_j - L_{\text{min}},
\end{equation}
As shown in Fig.~\ref{fig:PathLengthDrywallH}, the path length difference (expressed in wavelengths $\lambda$) increases with the off-specular angle, resulting in faster oscillations of the \gls{ssf}. This behavior confirms that the \gls{ssf} is not stationary with respect to the scattering angle, and therefore can not be reliably filtered using a fixed angular window.

\subsection{Window Size Selection}

The choice of the sliding-window length $W$ represents a trade-off between preserving the large-scale trend and suppressing small-scale fluctuations. To quantify this trade-off, we define the \gls{ssf} residual as
\begin{equation}
    e_W(k) = 10 \log_{10} \left(
    \frac{|\mathbf{E}_{\mathrm{S,CP/XP}}(k)|^2_{\mathrm{raw}}}
    {|\mathbf{E}_{\mathrm{S,CP/XP}}(k)|^2_{\mathrm{smoothed}(W)}}
    \right),
\end{equation}
where $k$ indexes the MUT orientation. The raw power is obtained from the coherent sum of MPCs, while the smoothed power is computed using a sliding window of length $W$.

Ideally, $e_W(k)$ should capture only the \gls{ssf} component after removal of the \gls{lst}. We evaluate the quality of the separation using the quantity $\sigma_e$, i.e., the standard deviation of $|e_W|$, expressed in dB units \cite{554753}.
The result shown in Fig.~\ref{fig:LSF_SSF_sigma_HH}, reveal three distinct regimes:
\begin{enumerate}
    \item $W < 5$: the smoothing window is too short and partially tracks the \gls{ssf}, leading to underestimation of its variability;
    \item $W = 5$: both metrics reach a plateau, indicating that the \gls{ssf} is fully captured in the residual while the \gls{lst} is preserved;
    \item $W > 5$: the window becomes too large, and part of the \gls{lst} is absorbed into the residual, leading to overestimation of \gls{ssf}.
\end{enumerate}
Based on this analysis, a window length of $W=5$ is selected for all materials. An example of the resulting large-scale trend extraction is shown in Fig.~\ref{fig:Carpet_HH_Raw_LSF}.

\bibliographystyle{IEEEtran}
\bibliography{main}

\end{document}

%% file: MyPackages.tex
\usepackage{amsmath,amssymb,mathtools}
\usepackage{bm}
\usepackage{physics}
\usepackage{booktabs}
\usepackage{tkz-euclide}
\usetikzlibrary{backgrounds}
\usepackage{textcomp}
\usepackage{refcount}
\usepackage{xfrac}
\usepackage{algorithm}
\usepackage{algorithmic}
\usetikzlibrary{arrows.meta, positioning, shapes.geometric, shapes.symbols, calc}
\usepackage{xcolor} 
\definecolor{rowgray}{gray}{0.93}
\usepackage{graphicx}
\usepackage{svg}
\svgsetup{inkscapepath=svg-inkscape/}
\usepackage{comment}
\usepackage[utf8]{inputenc}
\usepackage{textgreek}

\usepackage{glossaries}
\setacronymstyle{long-short}

\loadglsentries{MyAcronyms}

%% file: MyAcronyms.tex
\newacronym{er}{ER}{Effective Roughness}
\newacronym{ds}{DS}{Diffuse Scattering}
\newacronym{mpc}{MPC}{Multipath Components}
\newacronym{pg}{PG}{Path Gain}
\newacronym{sp}{SR}{Specular Reflection}
\newacronym{mut}{MUT}{Material Under Test}
\newacronym{em}{EM}{Electromagnetic}
\newacronym{AoA}{AoA}{Angle-of-Arrival}
\newacronym{sage}{SAGE}{Space-Alternating Generalized Expectation-maximization}
\newacronym{xpd}{XPD}{Cross-Polarization Discrimination}
\newacronym{ssa}{SSA}{Singular Spectrum Analysis}
\newacronym{rer}{RER}{Reciprocal ER}
\newacronym{g-rer}{G-RER}{Gaussian Reciprocal ER}
\newacronym{rms}{RMS}{Root Mean Square}
\newacronym{rmsd}{RMSD}{Root Mean Square Deviation}
\newacronym{rmse}{RMSE}{Root Mean Square Error}
\newacronym{go}{GO}{Geometrical Optics}
\newacronym{ka}{KA}{Kirchhoff Approximation}
\newacronym{iem}{IEM}{Integral Equation Method}
\newacronym{spm}{SPM}{Small Perturbation Method}
\newacronym{tx}{TX}{transmitter}
\newacronym{rx}{RX}{receiver}
\newacronym{rcs}{RCS}{Radar Cross Sections}
\newacronym{po}{PO}{Physical Optics}
\newacronym{pso}{PSO}{Particle Swarm Optimization}
\newacronym{h}{H}{horizontal}
\newacronym{v}{V}{vertical}
\newacronym{hh}{HH}{horizontal-horizontal}
\newacronym{hv}{HV}{horizontal-vertical}
\newacronym{vh}{VH}{vertical-horizontal}
\newacronym{vv}{VV}{vertical-vertical}
\newacronym{cp}{CP}{co-polarized}
\newacronym{xp}{XP}{cross-polarized}
\newacronym{ffz}{FFZ}{First Fresnel Zone}
\newacronym{aoa}{AoA}{angle-of-arrival}
\newacronym{dt}{DT}{digital twin}
\newacronym{rt}{RT}{ray tracing}
\newacronym{te}{TE}{Transverse Electric}
\newacronym{tm}{TM}{Transverse Magnetic}
\newacronym{lst}{LST}{large-scale trend}
\newacronym{ssf}{SSF}{small-scale fading}
\newacronym{los}{LoS}{Line-of-Sight}
\newacronym{gbsm}{GBSM}{Geometrical-Based Stochastic Modeling}
\newacronym{rf}{RF}{radio frequency}
\newacronym{nlos}{NLOS}{Non-Line-of-Sight}

%% file: Pictures/ReferenceSystem.tex
\usetikzlibrary{backgrounds,calc,decorations.pathmorphing}
\definecolor{volbrownA}{RGB}{222,196,156}
\definecolor{volbrownB}{RGB}{210,180,140}
\definecolor{volbrownC}{RGB}{198,164,120}
\definecolor{surfbrown}{RGB}{205,175,130}

\pgfkeys{
  /wavearc/.is family, /wavearc,
  color/.store in=\wacolor,     color=teal,
  Rstart/.store in=\waRstart,   Rstart=0.88,
  dR/.store in=\wadR,           dR=0.22,
  xscale/.store in=\waxscale,   xscale=1,
  yscale/.store in=\wayscale,   yscale=0.80,
  a1/.store in=\waa,            a1=18,
  a2/.store in=\wab,            a2=20,
  a3/.store in=\wac,            a3=22,
  lw1/.store in=\walwa,         lw1=0.4pt,
  lw2/.store in=\walwb,         lw2=0.7pt,
  lw3/.store in=\walwc,         lw3=1.1pt,
  label/.store in=\walabel,     label=,
  lx/.store in=\walx,           lx=0,
  ly/.store in=\waly,           ly=0
}

\newcommand{\WavefrontArcs}[2][]{%
  \begin{scope}[#1]
    \pgfkeys{/wavearc,
      color=teal,
      Rstart=0.88,
      dR=0.22,
      xscale=1,
      yscale=0.80,
      a1=18, a2=20, a3=22,
      lw1=0.4pt, lw2=0.7pt, lw3=1.1pt,
      label=, lx=0, ly=0
    }
    \pgfkeys{/wavearc,#2}

    \begin{scope}[xscale=\waxscale, yscale=\wayscale]
      \pgfmathsetmacro{\RA}{\waRstart}
      \draw[\wacolor, line width=\walwa, opacity=0.45]
        ({\RA*cos(-\waa)},{\RA*sin(-\waa)})
        arc[start angle=-\waa, end angle=\waa, radius=\RA];

      \pgfmathsetmacro{\RB}{\waRstart+\wadR}
      \draw[\wacolor, line width=\walwb, opacity=0.65]
        ({\RB*cos(-\wab)},{\RB*sin(-\wab)})
        arc[start angle=-\wab, end angle=\wab, radius=\RB];

      \pgfmathsetmacro{\RC}{\waRstart+2*\wadR}
      \draw[\wacolor, line width=\walwc, opacity=0.90]
        ({\RC*cos(-\wac)},{\RC*sin(-\wac)})
        arc[start angle=-\wac, end angle=\wac, radius=\RC];

      \node at (\walx,\waly) {\walabel};
    \end{scope}
  \end{scope}
}

\begin{tikzpicture}[scale=1.5]

  \pgfmathsetmacro{\r}{2.5}
  \pgfmathsetmacro{\thetaX}{195}
  \pgfmathsetmacro{\thetaY}{335}
  \pgfmathsetmacro{\thetaZ}{90}

  \pgfmathsetmacro{\cx}{cos(\thetaX)} \pgfmathsetmacro{\sx}{sin(\thetaX)}
  \pgfmathsetmacro{\cy}{cos(\thetaY)} \pgfmathsetmacro{\sy}{sin(\thetaY)}
  \pgfmathsetmacro{\cz}{cos(\thetaZ)} \pgfmathsetmacro{\sz}{sin(\thetaZ)}
  \pgfmathsetmacro{\Ramp}{1.00}

  \def\Xmin{-2.2} \def\Xmax{2.2}
  \def\Ymin{-1.8} \def\Ymax{1.8}
  \def\bandstep{0.30}
  \def\gridstep{0.40}

  \pgfmathdeclarefunction{seasurf}{2}{%
    \pgfmathparse{ -0.55
      + \Ramp*(%
          0.18*sin(deg(#1))*cos(deg(#2))
        + 0.12*sin(deg(2.1*#1 + 0.3*#2))
        + 0.15*sin(deg(3.7*#1 - 1.5*#2))
        + 0.10*sin(deg(6.0*#1 + 2.0*#2))
        + 0.06*sin(deg(9.0*#1 - 4.0*#2)))}%
  }

  \def\Zbase{-2.5}

  \begin{scope}[on background layer,
    x={(\cx*0.35cm,\sx*0.35cm)},
    y={(\cy*0.35cm,\sy*0.35cm)},
    z={(\cz*0.35cm,\sz*0.35cm)}]


    \path[fill=volbrownB!75!white, draw=brown!55!black, line width=0.2pt, opacity=0.85]
      plot[samples=80, domain=\Ymin:\Ymax, variable=\yy, smooth]
        ({\Xmin},{\yy},{seasurf(\Xmin,\yy)})
      -- ({\Xmin},{\Ymax},{\Zbase})
      -- ({\Xmin},{\Ymin},{\Zbase})
      -- cycle;

    \path[fill=volbrownA!85!white, draw=brown!55!black, line width=0.2pt, opacity=0.85]
      plot[samples=80, domain=\Xmin:\Xmax, variable=\xx, smooth]
        ({\xx},{\Ymax},{seasurf(\xx,\Ymax)})
      -- ({\Xmax},{\Ymax},{\Zbase})
      -- ({\Xmin},{\Ymax},{\Zbase})
      -- cycle;

    \path[fill=volbrownA!65!white, draw=brown!45!black, line width=0.2pt, opacity=0.70]
      ({\Xmin},{\Ymin},{\Zbase})
      -- ({\Xmax},{\Ymin},{\Zbase})
      -- ({\Xmax},{\Ymax},{\Zbase})
      -- ({\Xmin},{\Ymax},{\Zbase})
      -- cycle;


    \path[fill=volbrownC!70!white, draw=brown!60!black, line width=0.2pt, opacity=0.90]
      plot[samples=80, domain=\Xmin:\Xmax, variable=\xx, smooth]
        ({\xx},{\Ymin},{seasurf(\xx,\Ymin)})
      -- ({\Xmax},{\Ymin},{\Zbase})
      -- ({\Xmin},{\Ymin},{\Zbase})
      -- cycle;

    \path[fill=volbrownC!70!white, draw=brown!60!black, line width=0.2pt, opacity=0.90]
      plot[samples=80, domain=\Ymin:\Ymax, variable=\yy, smooth]
        ({\Xmax},{\yy},{seasurf(\Xmax,\yy)})
      -- ({\Xmax},{\Ymax},{\Zbase})
      -- ({\Xmax},{\Ymin},{\Zbase})
      -- cycle;

    \begin{scope}
      \clip
        plot[samples=120, domain=\Xmin:\Xmax, variable=\xx, smooth]
          ({\xx},{\Ymin},{seasurf(\xx,\Ymin)}) --
        plot[samples=120, domain=\Ymin:\Ymax, variable=\yy, smooth]
          ({\Xmax},{\yy},{seasurf(\Xmax,\yy)}) --
        plot[samples=120, domain=\Xmax:\Xmin, variable=\xx, smooth]
          ({\xx},{\Ymax},{seasurf(\xx,\Ymax)}) --
        plot[samples=120, domain=\Ymax:\Ymin, variable=\yy, smooth]
          ({\Xmin},{\yy},{seasurf(\Xmin,\yy)}) -- cycle;

      \foreach \yy in {-1.8,-1.5,-1.2,-0.9,-0.6,-0.3,0,0.3,0.6,0.9,1.2,1.5}{
        \pgfmathsetmacro{\yya}{\yy}
        \pgfmathsetmacro{\yyb}{\yy+\bandstep}
        \path[fill=surfbrown!70!white, opacity=0.45]
          plot[samples=90, domain=\Xmin:\Xmax, variable=\xx, smooth]
            ({\xx},{\yya},{seasurf(\xx,\yya)})
          --
          plot[samples=90, domain=\Xmax:\Xmin, variable=\xx, smooth]
            ({\xx},{\yyb},{seasurf(\xx,\yyb)})
          -- cycle;
      }

      \tikzset{surface grid/.style={line cap=round, line join=round,
               very thin, brown!55!black, opacity=.45}}

      \foreach \yy in {-1.8,-1.4,-1.0,-0.6,-0.2,0.2,0.6,1.0,1.4,1.8}{
        \draw[surface grid]
          plot[samples=140, domain=\Xmin:\Xmax, variable=\xx, smooth]
            ({\xx},{\yy},{seasurf(\xx,\yy)});
      }
      \foreach \xx in {-2.2,-1.8,-1.4,-1.0,-0.6,-0.2,0.2,0.6,1.0,1.4,1.8,2.2}{
        \draw[surface grid]
          plot[samples=140, domain=\Ymin:\Ymax, variable=\yy, smooth]
            ({\xx},{\yy},{seasurf(\xx,\yy)});
      }
    \end{scope}

  \end{scope}

  \draw[->, gray!70!black]
    (0,0) -- ({\r*cos(\thetaY)},{\r*sin(\thetaY)}) node[below right] {$y$};
  \draw[->, gray!70!black]
    (0,0) -- ({\r*cos(\thetaX)},{\r*sin(\thetaX)}) node[below left]  {$x$};
  \draw[->, gray!70!black]
    (0,0) -- ({\r*cos(\thetaZ)},{\r*sin(\thetaZ)}) node[above]       {$z$};

  \shade[shading=radial, inner color=violet!70, outer color=violet!25,
         rotate=-45, scale=0.38, opacity=0.75]
    (0,0) .. controls (-4,6) and (4,6) .. (0,0) -- cycle;

  \draw[->, thick, gray!70!black] (0,0) ++(130:1.2cm) arc (130:95:1.2cm)
    node[midway, sloped, above] {$\vartheta_i$};
  \draw[->, thick, gray!70!black] (0,0) ++(75:1.2cm) arc (75:85:1.2cm)
    node[midway, sloped, above] {$\vartheta_s$};
  \draw[->, thick, gray!70!black] (0,0) ++({\thetaX+10}:0.3cm) arc ({\thetaX+10}:340:0.3cm)
    node[pos=0.55, sloped, below] {$\varphi_i$};
  \draw[->, thick, gray!70!black] (0,0) ++({\thetaX+5}:0.7cm) arc ({\thetaX+5}:420:0.7cm)
    node[pos=0.45, sloped, below] {$\varphi_s$};

  \WavefrontArcs[
    shift={(-2.70,2.70)},
    rotate=-45
  ]{color=blue!75!black, Rstart=0.88, xscale=1, yscale=0.80,
    a1=18, a2=20, a3=22,
    lw1=0.5pt, lw2=0.8pt, lw3=1.2pt}

  \WavefrontArcs[
    shift={(0.75,0.75)},
    rotate=45
  ]{color=green!60!black, Rstart=0.88, xscale=1, yscale=0.80,
    a1=18, a2=20, a3=22,
    lw1=0.5pt, lw2=0.8pt, lw3=1.2pt,
    label={$\mathbf{E_r}$}, lx=1.0, ly=-0.70}

  \WavefrontArcs[
    shift={(0.05,-0.25)},
    rotate=-45
  ]{color=orange!85!black, Rstart=0.88, xscale=1, yscale=0.80,
    a1=18, a2=20, a3=22,
    lw1=0.5pt, lw2=0.8pt, lw3=1.2pt,
    label={$\mathbf{E_t}$}, lx=1.0, ly=-0.55}


  \draw[dotted, gray!60] (2,2) -- (2,1);
  \draw[dotted, gray!60] (2,1) -- (0,0);
  \draw[dotted, gray!60] (0,0) -- (2,2);
  \draw[->, very thick, green!60!black] (1.45,1.45) -- (2,2);
  \filldraw[green!60!black] (1.45,1.45) circle (0.8pt);
  \node[above right, xshift=-20pt, yshift=-5pt] at (2,2) {$\widehat{\mathbf{k}}_r$};

  \draw[dotted, gray!60] (-2,2) -- (-2,0.5);
  \draw[dotted, gray!60] (-2,0.5) -- (2,-0.5);
  \draw[dotted, gray!60] (0,0) -- (-2,2);
  \draw[->, very thick, blue!75!black] (-2,2) -- (-1.5,1.5);
  \filldraw[blue!75!black] (-2,2) circle (0.8pt);
  \node[above right, xshift=8pt] at (-1.7,1.5) {$\widehat{\mathbf{k}}_i$};
  \node[above right, xshift=0pt, yshift=-30pt] at (-2,2) {$\mathbf{E_i}$};

  \draw[dotted, gray!60] (0,0) -- (0.9,-1.1);
  \draw[->, very thick, orange!85!black] (0.77,-0.95) -- (1.1687,-1.4284);
  \filldraw[orange!85!black] (0.77,-0.95) circle (0.8pt);
  \node[below right, xshift=10pt, yshift=0pt] at (0.89,-1.09) {$\widehat{\mathbf{k}}_t$};

  \draw[dotted, gray!60] (0,0) -- (0.9674,2.6579);
  \draw[->, very thick, violet!80!black] (0.7,1.92) -- (0.9674,2.6579);
  \filldraw[violet!80!black] (0.7,1.92) circle (0.8pt);
  \node[above left, xshift=-5pt, yshift=-9pt] at (0.9674,2.6579) {$\widehat{\mathbf{k}}_s$};

  \node[above left, xshift=-12pt, yshift=-35pt] at (1.6,1.6) {$\mathbf{E_s}$};

  \node at (-1.0,0.2) {\textbf{dS}};

  \draw[<->, thick, gray!70!black] (0,0) ++(47:1.2cm) arc (45:66:1.2cm)
    node[midway, sloped, above] {$\psi$};

\end{tikzpicture}

%% file: Pictures/Testbed.tex
\usetikzlibrary{calc,decorations.pathreplacing,arrows.meta}

\newcommand{\ROT}[1][rotate=0]{%
    \tikz [x=0.25cm,y=0.60cm,line width=.3ex,-stealth,#1] \draw (0,0) arc (-150:150:1 and 1);%
}

\newcommand{\TILT}[1][rotate=0]{%
    \tikz [x=0.25cm,y=0.60cm,line width=.3ex,-stealth,#1] \draw (0,0) arc (35:-35:1 and 1);%
}

\begin{tikzpicture}[line cap=round, line join=round,
  dot/.style={circle,fill,inner sep=1.7pt}, scale=0.6]
  \path[use as bounding box] (-5.5,-5.0) rectangle (13.2,10.2);

\begin{scope}[xshift=0cm,yshift=0cm]


\coordinate (WBL) at (1.0,0.5);    
\coordinate (WBR) at (11.0,-0.2);  
\coordinate (WTL) at (1.0,9.5);    
\coordinate (WTR) at (11.0,8.8);   

\draw[thick, fill=brown, opacity=0.8] (WBL) -- (WTL) -- (WTR) -- (WBR) -- cycle;

\begin{scope}
  \clip (WBL) -- (WTL) -- (WTR) -- (WBR) -- cycle;
  \pgfmathtruncatemacro{\N}{10}
  \pgfmathtruncatemacro{\Nm}{\N-1}
  \foreach \k in {1,...,\Nm}{
    \pgfmathsetmacro{\t}{\k/\N}
    \coordinate (P) at ($(WBL)!\t!(WTL)$);
    \coordinate (Q) at ($(WBR)!\t!(WTR)$);
    \draw[very thin, white, opacity=.35] (P) -- (Q);
  }
  \foreach \k in {1,...,\Nm}{
    \pgfmathsetmacro{\t}{\k/\N}
    \coordinate (P) at ($(WBL)!\t!(WBR)$);
    \coordinate (Q) at ($(WTL)!\t!(WTR)$);
    \draw[very thin, white, opacity=.35] (P) -- (Q);
  }
\end{scope}

\draw[<->] ($(WTL)+(0.075,.5)$) -- node[above] {$L_{\scriptscriptstyle \text{MUT}}$} ($(WTR)+(-.075,.5)$);
\draw[<->] ($(WTR)+(.5,0.075)$) -- node[right] {$L_{\scriptscriptstyle \text{MUT}}$} ($(WBR)+(.5,-.075)$);

\coordinate (PMUT) at ($(WBL)!0.5!(WTR)+(0,-0.5)$);
\node[dot,label={[xshift=0pt,yshift=10pt]below left:$\mathbf{P}_{\scriptscriptstyle \text{MUT}}$}] at (PMUT) {};

\draw[-{Stealth[length=2.5mm]}] (PMUT) -- ++(-0.11,-1.5)
  node[left,pos=.95] {$\hat{\mathbf{n}}_{\scriptscriptstyle \text{MUT}}$};

\draw[-{Stealth[length=2.5mm]}] (PMUT) -- ++(0.8,-0.06)
  node[below, xshift=5pt, yshift=0pt, font=\scriptsize] {$\text{X}_{\text{MUT}}$};
\draw[-{Stealth[length=2.5mm]}] (PMUT) -- ++(0.0,0.8)
  node[right, xshift=-9pt, yshift=4pt, font=\scriptsize] {$\text{Y}_{\scriptscriptstyle \text{MUT}}$};

\coordinate (PTR) at ($(PMUT)+(2.8,-0.2)$);
\draw[solid, draw opacity=0.3] (PMUT) -- (PTR) node[midway, above, yshift=-3pt] {$d$};
\node[dot,label={[xshift=0pt,yshift=2pt]right:$\mathbf{P}_{\scriptscriptstyle \text{T/R}}$}] at (PTR) {};

\coordinate (DSN) at ($(PMUT)+(-3.5,1.2)$);
\node[dot,scale=0.5,fill=black, label={[xshift=0pt,yshift=2pt]right:$\mathrm{dS_n}$}] at (DSN) {};

\coordinate (PSR) at ($(PMUT)+(0.8,3.0)$);
\coordinate (PJ) at ($(PSR)+(-0.3,-0.6)$);

\node[inner sep=0pt] at (PSR) {\textcolor{black}{\Large$\times$}};

\draw[black, thick] (PSR) ellipse [x radius=1.1cm, y radius=1.0cm];

\draw[<->, black] ($(PSR)+(0,.05)$) -- node[left] {$\rho$} ($(PSR)+(0,0.95cm)$);

\node[dot,scale=0.5,fill=red, label={[xshift=-3pt,yshift=0pt,text=red]right:$\mathbf{p}_\mathrm{j}$}] at (PJ) {};

\coordinate (TXbase) at (-3.5,-5.0);
\coordinate (TX)     at (-3.5,-2.5);

\node[inner sep=0pt] (RXimg) at (TX) {\includegraphics[width=3.5cm]{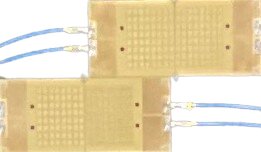}};

\draw [decorate,decoration={brace,amplitude=5pt,raise=3pt}]
  ($(TXbase)+(0.0,0.15)$) -- ($(TX)+(0.0,-0.15)$) node[midway, xshift=-20pt,yshift=-2pt]{$h_{\scriptscriptstyle \text{RX}}$};

\draw (TXbase) -- ($(TX)+(0,0)$);
\draw ($(TXbase)+(-.25,0)$) -- ($(TXbase)+(.25,0)$);

\coordinate (RXbase) at (11.0,-4.5);
\coordinate (RX)     at (11.0,-2.5);

\node[inner sep=0pt] (TXimg) at (RX) {\includegraphics[width=2.0cm]{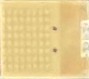}};

\draw [decorate,decoration={brace,amplitude=5pt,raise=3pt}]
  ($(RXbase)+(0.0,0.15)$) -- ($(RX)+(0.0,-0.15)$) node[midway, xshift=-20pt,yshift=2pt]{$h_{\scriptscriptstyle \text{TX}}$};

\draw (RXbase) -- ($(RX)+(0,0)$);
\draw ($(RXbase)+(-.25,0)$) -- ($(RXbase)+(.25,0)$);

\draw (TX) -- (PMUT) node[pos=0.4,sloped,below] {$L_{\scriptscriptstyle \text{RX}}$};
\draw (RX) -- (PMUT) node[pos=0.70,sloped,below] {$L_{\scriptscriptstyle \text{TX}}$};

\draw[dashed] (TX) -- (DSN) node[pos=0.45,sloped,above] {$r_{\scriptscriptstyle i,n}$};
\draw[dashed] (RX) -- (DSN) node[pos=0.5,sloped,below] {$r_{\scriptscriptstyle s,n}$};

\draw[dashed] (TX) -- (PSR) node[pos=0.4,sloped,above] {$r_{\scriptscriptstyle i,\textbf{SR}}$};
\draw[dashed] (RX) -- (PSR) node[pos=0.3,sloped,above] {$r_{\scriptscriptstyle s,\textbf{SR}}$};

\draw[red] (TX) -- (PJ) node[pos=0.55,sloped,below,text=red] {$r_{\scriptscriptstyle i,\ell}$};
\draw[red] (RX) -- (PJ) node[pos=0.6,sloped,below,text=red] {$r_{\scriptscriptstyle s,\ell}$};

\coordinate (MUTbase) at ($(PTR)+(0,-4.5)$);
\coordinate (MUTtop)  at ($(PTR)+(0,0)$);
\draw[solid] (MUTbase) -- ($(MUTbase)+(0,0.5)$);
\draw[solid, draw opacity=0.3] (MUTbase) -- (MUTtop);
\draw ($(MUTbase)+(-.25,0)$) -- ($(MUTbase)+(.25,0)$);

\draw [decorate,decoration={brace,amplitude=5pt,mirror,raise=3pt}]
  ($(MUTbase)+(0.0,0.15)$) -- ($(MUTtop)+(0.0,-0.15)$) node[midway, xshift=20pt, yshift=2pt]{$h_{\scriptscriptstyle \text{MUT}}$};

\draw[dotted] (PTR) -- ($(PTR)+(0,6.9)$);
\draw ($(PTR)+(0,6.0)$) node {\color{green!50!black}\ROT[rotate=-90]} node[right,xshift=-14pt] {\color{green!50!black}ROT};

\draw[dotted] (PMUT) -- ++(-6.9,0.35);
\path (PMUT) -- ++(-5.9,0.35) node (tilt) {\color{green!50!black}\TILT[rotate=174]} node[below, yshift=-8pt] {\color{green!50!black}TILT};

\node[above right, xshift=0pt, yshift=0pt] at (TXimg.north east) {};
\node[dot,label={[xshift=0pt,yshift=2pt]right:$\mathbf{P}_{\scriptscriptstyle \text{TX}}$}] at (TXimg) {};

\node[above left, xshift=0pt, yshift=0pt] at (RXimg.north east) {};
\node[dot,label={[xshift=-30pt,yshift=2pt]right:$\mathbf{P}_{\scriptscriptstyle \text{RX}}$}] at (RXimg) {};

\end{scope}
\end{tikzpicture}